\documentstyle[preprint,aps]{revtex}
\begin{document}
\newtheorem{theorem}{Theorem}
\newtheorem{defin}{Definition}
\newtheorem{prop}{Proposition}
\draft
\title{Nearly Autoparallel Maps, Tensor Integral and\\
   Conservation Laws on Locally Anisotropic Spaces}
\author{Sergiu I. Vacaru}
\address{Institute of Applied Physics,Academy of Sciences of Moldova,\\
 5 Academy str., Chi\c sin\v au 277028, Republic of Moldova }
\date{\today}
\maketitle
\begin{abstract}
We formulate the theory of nearly autoparallel maps (generalizing conformal
transforms) of locally anisotropic spaces and define the nearly
 autoparallel integration as the inverse operation to both covariant derivation
and deformation of connections by nearly autoparallel maps. By using this
geometric formalism we consider a variant of solution of the problem of
formulation of conservation laws for locally anisotropic gravity. We note
that locally anisotropic spases contain as particular cases various
 extensions of Kaluza--Klein, generalized Lagrange and Finsler spaces.

{\bf \copyright \ S.I.Vacaru}
\end{abstract}

\pacs{04.50.+h, 02.40.+k, 04.20.Cv, 04.90.+e}

\widetext
\section{ Introduction}
Theories of field  interactions on locally anisotropic
curved spaces form a new branch of modern theoretical and mathematical
physics. They are used for modelling in a self--consistent manner physical
processes in locally anisotropic, stochastic and turbulent media
with beak radiational reaction and diffusion [1-8]. The first model of locally
anisotropic space was proposed by P.Finsler [9] as a generalization of
 Riemannian geometry; here we also cite the fundamental contribution made
 by E. Cartan [10] and mention that in monographs [5-7]  detailed
 bibliographies are contained. In this work we follow conventions of R.Miron
and M.Anastasiei [1,2] and base our investigations on their general model
of locally anisotropic (la) gravity (in brief we shall write la-gravity)
on vector bundles, v--bundles, provided with nonlinear and distinguished
connection and metric structures (we call a such type of v--bundle as a
la-space if connections and metric are compatible).

The study  of models of classical and quantum field interactions
on la-spaces is in order of the day. For instance, in papers [11,12]
the problem of definition of spinors on la-spaces is solved and some
models of locally anisotropic Yang--Mills and gauge like gravitational
interactions are analyzed (see  alternative approaches in [5,8]).
The development of this direction entails great difficulties because
of problematical character of the possibility and manner of definition
of conservation laws on la-spaces. It will be recalled that, for instance,
conservation laws of energy--momentum type are a consequence of existence
of a global group of automorphisms of the fundamental Mikowski spaces
(for (pseudo)Riemannian spaces one considers automorphisms on tangent
bundle and particular cases when there are symmetries generated by
existence of Killing vectors). No global or local automorphisms exist on
generic la-spaces and in result of this fact the formulation of
la-conservation laws is sophisticated and full of ambiguities. R. Miron
and M. Anastasiei  firstly pointed out the nonzero divergence of the matter
energy-momentum tensor, the source in Einstein equations on la-spaces, and
considered an original approach to the geometry of time--dependent Lagrangians
[1,2,13].  Nevertheless, the rigorous definition of energy-momentum values
for la-gravitational and matter fields and the form of conservation laws
for such values have not been considered in present--day studies of the
mentioned problem.

Our aim  is to develop a necessary geometric background
(the theory of nearly autoparallel maps and tensor integral formalism on
la-multispaces) for formulation and a detailed investigation of conservation laws
on la-spaces.

The question of definition of tensor integration as the inverse operation of
covariant derivation was posed and studied by A.Mo\'or [14]. Tensor--integral
and bitensor formalisms turned out to be very useful in solving certain
 problems connected with conservation laws in general relativity [15,16].
 In order to extend tensor--integral constructions we have proposed [17,18] to
 take into consideration nearly autoparallel [19-21] and nearly geodesic [22]
 maps, na-- and ng--maps, which forms a subclass of local 1--1 maps of
 curved spaces with deformation of the connection and metric structures.
A generalization of the Sinyukov's ng--theory for spaces with local anisotropy
was proposed by considering maps with deformation of connection for
Lagrange spaces (on Lagrange spaces see [23,1,2]) and
generalized Lagrange spaces [24-27]. Tensor integration formalism for
 generalized Lagrange spaces was developed in [28,29]. One of the main
purposes of this paper is to synthesize the results obtained in the mentioned
 works and  to formulate them for a very general class of la--spaces.
As the next step the la--gravity and analysis of la--conservation laws
are considered.

We note  that proofs of our theorems are mechanical, but, in most cases,
they are rather tedious calculations similar to those presented in [22,20,30].
We sketch  some of them in  Appendixes A and B.

In Sec. II we present some basic results, necessary for our further
considerations, on nonlinear connections in bundle spaces [1,2]. Sec. III is
 devoted to the formulation of the theory of nearly autoparallel maps of
 la--spaces.  Classification of na--maps and formulation of their invariant
conditions are given in Sec. IV . In Sec. V we define the nearly
 autoparallel tensor--integral on locally anisotropic multispaces. The problem
of formulation of conservation laws on spaces with local anisotropy is
studied in Sec. VI . We present a definition of conservation laws for
la--gravitational fields on na--images of la--spaces in Sec. VII .
Outlook and conclusions are contained in Sec. VIII .

\section{ Geometry of Locally Anisotropic Spaces }

As a general model of locally anisotropic space (generalizing the concept of
Finsler and Lagrange spaces, see R.Miron and M.Anastasiei monographs [1,2])
we consider a (for simplicity, trivial) v--bundle
 ${\xi} = (E,F,{\pi}, M),$ where $M$ is a
 differentiable manifold of dimension $n, dim M =n,$ the typical fibre $F$ is a
real vector space of dimension $m$ and ${\pi}: E \to M $ is a differentiable
surjection. Local coordinates on the base space $M$ are denoted by
 $x = ( x^i )$ and on the total space $E$ by
$u=(x,y)={u^{\alpha}}=(x^{i},y^{a})$ (we shall use indices $i,j,k,..=1,2,...n$
for $M$-components, $a,b,c,...=1,2,...m$ for $F$-components, $y$-coordinates
on $F$ can be interpreted  as parameters of local anisotropy, and Greek indices
$\alpha , \beta ,...$ as general ones on v-bundle $\xi ).$

\begin{defin}
 A nonlinear connection, briefly an N-connection,
 in tangent bundle
${\xi}$  is a global decomposition of the tangent bundle $T{\xi}$ into
horizontal $H{\xi}$  and vertical $V{\xi}$  subbundles :
\begin{equation}
T{\xi}=H{\xi} {\oplus} V{\xi}. \label {f1}
\end{equation}
\end{defin}

To an N--connection on ${\xi}$~ one can associate a covariant derivation on
$M$:
\begin{equation}
{\nabla}_{Y}A={Y^{i}}{\lbrace}{{{\partial}A^{a}}\over{{\partial}x^{i}}}+
N^{a}_{i}(x,A){\rbrace}s_{a}, \label{f2}
\end{equation}
where $s_{a}$ are local linearly independent sections of ${\xi},$
$A=A^{a}s_{a}$  and $Y=Y^{i}s_{i}$ are vector fields. The differentiable
 functions $N^{a}_{i} ,$ as functions of $x^i$ and $y^i$, i.e.
$N^{a}_{i}(x,y),$ are called the coefficients of the N--connection.

In v--bundle ${\xi}$ we can define a local basis (frame) adapted
to a given N--connection :
\begin{equation}
X_{\alpha}=(X_{i}={{\delta}\over{{\delta}x^{i}}}=
{\partial}_{i}-N^{a}_{i}(x,y){{\partial}\over{{\partial}y^{a}}}, \,
X_{a}={{\delta}\over{{\delta}y^{a}}}=
      {{\partial}\over{{\partial}y^{a}}}) \label{f3}
\end{equation}
The dual basis to $X_{\alpha}$ is written as
$$X^{{\alpha}}=(X^{i}={{\delta}x^{i}}=dx^{i}, \ ,
X^{a}={{\delta}y^{a}}=dy^{a}+N^{a}_{i}(x,y)dx^{i}). $$

By using  adapted bases  one can introduce the algebra of distinguished
tensor  fields (d-fields or d-tensors)
${\cal T}={\bigoplus}{\cal T}^{pr}_{qs}$~ on ${\xi},$
 which is the tensor algebra on the vector bundle (v--bundle) ${\xi}_{d}$
 defined by ${\pi}_{d} : H\xi {\bigoplus} V\xi {\to} \xi .$ An element $t {\in}
{\cal T}_{qs}^{pr},$  d--tensor field of type
$\left( \begin{array}{cc} p & r \\
			  q & s \end{array} \right) ,$
 is written  in local form as
$$  t=t^{{i}_{1}{\cdots}{i}_{p}{\,}{a}_{1}{\cdots}{a}_{r}}_
        {{j}_{1}{\cdots}{j}_{q}{\,}{b}_{1}{\cdots}{b}_{s}}(x,y)
{{\delta}\over{{\delta}x^{{i}_{1}}}}{\otimes}\,{\cdots}\,{\otimes}
{{\delta}\over{{\delta}x^{{i}_{q}}}}{\otimes}dx^{{j}_{1}}{\otimes}
\,{\cdots}\,dx^{{j}_{q}}{\otimes}$$
$${{\partial}\over{{\partial}y^{{a}_{1}}}}{\otimes}\,{\cdots}\,{\otimes}
{{\partial}\over{{\partial}y^{{a}_{r}}}}
{\otimes}{\delta}y^{{b}_{1}}
{\otimes}\,{\cdots}\,{\otimes}{\delta}y^{{b}_{s}}.$$

In addition  to d--tensors we can consider d--objects with various group
and coordinate transforms adapted to the global splitting (1).
\begin{defin}
 A linear d--connection is a linear connection $^{N}D$~ on
${\xi}$ conserving at a parallel translation the Whitney sum $H\xi {\bigoplus}
V\xi $ associated to a fixed N--connection on ${\xi}$.
\end{defin}

The components ${}^N{\Gamma}^{\alpha}_{{.}{\beta}{\gamma}}$ of d--connection
${}^{N}D$ are defined in the form :
$${}^ND_{\gamma}X_{\beta}  = {}^N{D_{X_{\gamma}}}X_{\beta} =
{}^N{\Gamma}^{\alpha}_{{.}{\beta}{\gamma}}X_{\alpha}$$

The torsion ${}^{N}T^{\alpha}_{{.}{\beta}{\gamma}}$ and  curvature
        ${}^{N}R^{{.}{\alpha}}_{{\beta}{.}{\gamma}{\delta}}$ of the connection
        ${}^{N}{\Gamma}^{\alpha}_{{.}{\beta}{\gamma}}$
can be introduced in a standard manner :
$$
{}^{N}T(X_{\gamma},X_{\beta}) = {}^N{T^{\alpha}_{{.}{\beta}{\gamma}}}
X_{\alpha},$$
where
\begin{equation}
{}^N{T^{\alpha}_{{.}{\beta}{\gamma}}} =
{}^N{{\Gamma}^{\alpha}_{{.}{\beta}{\gamma}}} -
{}^N{{\Gamma}^{\alpha}_{{.}{\gamma}{\beta}}} +
           w^{\alpha}_{{.}{\beta}{\gamma}},\label{f4}
\end{equation}

and $${}^N{R(X_{\delta},X_{\gamma},X_{\beta})} =
{}^N{R^{{.}{\alpha}}_{{\beta}{.}{\gamma}{\delta}}}
X_{\alpha},$$
where

\widetext
\begin{eqnarray}
{{}^N}{R^{\cdot \alpha}_{\beta \cdot \gamma \delta}} =
X_{\delta}{}^N{{\Gamma}^{\alpha}_{\cdot \beta \gamma}} -
X_{\gamma}{}^N{{\Gamma}^{\alpha}_{\cdot \beta \delta}} +
{}^N{{\Gamma}^{\varphi}_{\cdot \beta \gamma}}{}^N{{\Gamma}^{\alpha}
_{\cdot \varphi \delta}} -
 {}^N{{\Gamma}^{\varphi}_{\cdot \beta \delta}}{}^N{{\Gamma}^{\alpha}
_{\cdot \varphi \gamma}} +
{}^N{{\Gamma}^{\alpha}_{\cdot \beta \gamma}} {w^{\varphi}_{\cdot \gamma \delta}}.
\label{f5}
\end{eqnarray}
In formulas (4) and (5) we have used the nonholonomical coefficients $w^{\alpha}_
{{.}{\beta}{\gamma}}$~ of the adapted frame (3) defined as
\begin{eqnarray}
{\lbrack}X_{\alpha},X_{\beta}{\rbrack} = X_{\alpha}X_{\beta} -
                                       X_{\beta}X_{\alpha} =
w^{\gamma}_{{.}{\alpha}{\beta}}X_{\gamma}. \label {f6}
\end{eqnarray}

Now we introduce the metric (fundamental) tensor $g_{ij}(x,y).$ It is a second order,
covariant and nondegenerate d--tensor field on $M$~ (in general
                                             $g_{ij}(x,y)$~ is nonhomogeneous
in the variables $y^{i}$~ and it does not need to be generated by a fundamental
function and by a Lagrangian,  as in Finsler and, respectively, Lagrange geometry) .
For our purposes we consider v--bundles provided with a metric structure
\begin{equation}
G(u) = G_{\alpha \beta} du^{\alpha} du^{\beta} \label {f7}
\end{equation}
being compatible to given N--connection ${N^a_i}(u)$ and d--connection
$D$ structures. In this case we can split the metric (7) into horizontal
and vertical parts with respect to the locally adapted frame (3):
$$
 G=G_{{\alpha}{\beta}}{\delta}u^{\alpha}{\otimes}{\delta}u^{\beta} = $$
\begin{equation}
g_{ij}(x,y)dx^{i}{\otimes}dx^{j}+
h_{ab}(x,y){\delta}y^{a}{\otimes}{\delta}y^{b} \label {f8}
\end{equation}
with components of d--metric $h_{ab}(x,y)$~ defined from relations
\begin{equation}
 G_{ia}-N^{k}_{i}h_{ka} = 0. \label {f9}  \end{equation}
The compatibility conditions with d--connection $D$ are written as
\begin{equation}
 D_{\alpha}G_{{\beta}{\gamma}}=0. \label {f10}
\end{equation}
The components of a d--connection $D$ satisfying conditions (10) are denoted
as ${{\Gamma}^{\alpha}}_{\beta \gamma} .$ The corresponding formulas for
torsion ${T^{\alpha}}_{\beta \gamma}$ and
 curvature $R^{\cdot \alpha}_{\beta \cdot \gamma \delta}$
are written in a similar manner as (4) and (5) by omitting the left upper
index N.

We note that considering a similar to (10) metric structure on the tangent bundle
$TM$  being compatible with both N--connection and almost Hermitian structure
we obtain the so--called almost Hermitian model of generalized Lagrange geometry
[1,2], which contains as particular cases the Lagrange and Finsler geometry.
 Finally, we remark that la--geometric constructions
with respect to locally adapted frames are very similar to those for Riemannian
spaces with corresponding generalizations for torsion and nonmetricity.

\section{ Nearly Autoparallel Maps of Locally Anisotropic Spaces}

In this section we shall extend the ng--  [9] and  na--map [16,4-8,12,13]
theory  by introducing into consideration maps of vector bundles provided
with compatible N--connection, d--connection and metric structures.

Our geometric arena consists from pairs of open regions $( U,
{\underline  U})$ of la--spaces,
$ U{\subset}{\xi},\, {\underline U}{\subset}{\underline {\xi}}$, and 1--1
local maps $f :  U{\to}{\underline U}$  given by functions $f^{a}(u)$
of smoothly class $C^r( U) \, (r>2, $  or $r={\omega}$~ for analytic
functions) and their inverse functions $f^{\underline a}({\underline u})$
with corresponding non--zero Jacobians in every point $u{\in} U$ and
${\underline u}{\in}{\underline U}.$

We consider that two open regions $U$~ and ${\underline U}$~ are
attributed to a common for f--map coordinate system if this map is realized
on the principle of coordinate equality $q(u^{\alpha}) {\to} {\underline q}
(u^{\alpha})$~ for every point $q {\in} U$~ and its f--image ${\underline q}
{\in} {\underline U}.$  We note that all calculations included in this
work will be local in nature and taken to refer to open subsets of mappings
of type
$ {\xi} {\supset} U \buildrel f \over \longrightarrow {\underline U} {\subset}
{\underline {\xi}}.$
For simplicity, we suppose that in a fixed common coordinate system for
$U$ and ${\underline U}$ spaces $\xi$ and ${\underline {\xi}}$ are
characterized by a common N--connection structure (in consequence of (8) by
a corresponding concordance of d--metric structure), i.e.
$$N^{a}_{j}(u)={\underline N}^{a}_{j}(u)={\underline N}^{a}_{j}
({\underline u}),$$
which leads to the possibility to establish common  local bases, adapted to
a given N--connection, on both regions
$U$ and ${\underline U.}$  We consider that on $\xi$ it is defined the linear
d--connection structure with components
${\Gamma}^{\alpha}_{{.}{\beta}{\gamma}}.$
On the space $\underline {\xi}$ the linear d--connection is considered to be a
 general one with torsion
$$ {\underline T}^{\alpha}_{{.}{\beta}{\gamma}}={\underline
{\Gamma}}^{\alpha}_
{{.}{\beta}{\gamma}}-{\underline {\Gamma}}^{\alpha}_{{.}{\gamma}{\beta}}+
w^{\alpha}_{{.}{\beta}{\gamma}}.$$
and nonmetricity
\begin{equation}
 {\underline K}_{{\alpha}{\beta}{\gamma}}={{\underline D}_{\alpha}}
{\underline G}_{{\beta}{\gamma}}.  \label {f11}
\end{equation}

Geometrical objects on ${\underline {\xi}}$ are specified by underlined symbols
(for example, ${\underline A}^{\alpha}, {\underline
B}^{{\alpha}{\beta}})$~
or underlined indices (for example, $A^{\underline a}, B^{{\underline
a}{\underline b}}).$

For our purposes it is convenient to introduce auxiliary symmetric
d--connections,
${\gamma}^{\alpha}_{{.}{\beta}{\gamma}}={\gamma}^{\alpha}_{{.}{\gamma}{\beta}}
$~
on $\xi$ and ${\underline {\gamma}}^{\alpha}_{.{\beta}{\gamma}}=
                 {\underline {\gamma}}^{\alpha}_{{.}{\gamma}{\beta}}$ on
${\underline {\xi}}$  defined, correspondingly, as
$$ {\Gamma}^{\alpha}_{{.}{\beta}{\gamma}}=
              {\gamma}^{\alpha}_{{.}{\beta}{\gamma}}+
                   T^{\alpha}_{{.}{\beta}{\gamma}}\quad  {\rm\; and}\quad
  {\underline {\Gamma}}^{\alpha}_{{.}{\beta}{\gamma}}=
      {\underline {\gamma}}^{\alpha}_{{.}{\beta}{\gamma}}+
             {\underline T}^{\alpha}_{{.}{\beta}{\gamma}}.$$

We are interested in definition of local 1--1 maps from $U$ to ${\underline
U}$
characterized by symmetric, $P^{\alpha}_{{.}{\beta}{\gamma}},$ and
antisymmetric,
$Q^{\alpha}_{{.}{\beta}{\gamma}}$,~ deformations:
\begin{equation}
 {\underline {\gamma}}^{\alpha}_{{.}{\beta}{\gamma}}
={\gamma}^{\alpha}_{{.}{\beta}{\gamma}}+
P^{\alpha}_{{.}{\beta}{\gamma}}  \label {f12}
\end{equation}
and
\begin{equation}
 {\underline T}^{\alpha}_{{.}{\beta}{\gamma}}=
T^{\alpha}_{{.}{\beta}{\gamma}}+
Q^{\alpha}_{{.}{\beta}{\gamma}}.   \label {f13}
\end{equation}
The auxiliary linear covariant derivations induced by
${\gamma}^{\alpha}_{{.}{\beta}{\gamma}}$
and ${\underline {\gamma}}^{\alpha}_{{.}{\beta}{\gamma}}$~ are denoted
respectively
as $^{({\gamma})}D$~ and $^{({\gamma})}{\underline D}.$~

  Let introduce this local
coordinate parametrization of curves on $U$~:
$$u^{\alpha}=u^{\alpha}({\eta})=(x^{i}({\eta}),
y^{i}({\eta}),~{\eta}_{1}<{\eta}<{\eta}_{2},$$
where corresponding tangent vector field is defined as
$$  v^{\alpha}={{du^{\alpha}} \over d{\eta}}=
({{dx^{i}({\eta})} \over {d{\eta}}},
{{dy^{j}({\eta})} \over d{\eta}}).$$

\begin{defin}
 Curve $l$~ is called auto parallel, a--parallel, on
$\xi$  if its tangent vector field $v^{\alpha}$~ satisfies a--parallel
equations :
\begin{equation}
vDv^{\alpha}=v^{\beta}{^{({\gamma})}D}_{\beta}v^{\alpha}=
{\rho}({\eta})v^{\alpha},  \label {f14}
\end{equation}
where ${\rho}({\eta})$~ is a scalar function on $\xi$.
\end{defin}
Let curve ${\underline l} {\subset} {\underline {\xi}}$ is given  in parametric form
as $u^{\alpha}=u^{\alpha}({\eta}),~{\eta}_1 < {\eta} <{\eta}_2$
with tangent  vector field $v^{\alpha} = {{du^{\alpha}} \over {d{\eta}}}
{\ne} 0.$
We suppose that a 2--dimensional distribution $E_2({\underline l})$ is
defined along
${\underline l} ,$ i.e. in every point
$u {\in} {\underline l}$ is fixed a 2-dimensional vector space
$E_{2}({\underline l}) {\subset} {\underline {\xi}}.$  The introduced distribution
$E_{2}({\underline l})$~ is coplanar along ${\underline l}$~ if every vector
${\underline p}^{\alpha}(u^{b}_{(0)}) {\subset} E_{2}({\underline l}),
u^{\beta}_{(0)} {\subset} {\underline l}$~ rests contained in the same
distribution after parallel transports along ${\underline l},$~ i.e.
${\underline p}^{\alpha}(u^{\beta}({\eta})) {\subset} E_{2} ({\underline l}).$
\begin{defin}
 A curve ${\underline l}$~ is called nearly autoparallel, or in brief an
na--parallel, on space ${\underline {\xi}}$~ if a coplanar along
${\underline l}$~ distribution $E_{2}({\underline l})$  containing tangent to
${\underline l}$~ vector field $v^{\alpha}({\eta})$,~ i.e.
 $v^{\alpha}({\eta}) {\subset} E_{2}({\underline l}), $~ is defined.
\end{defin}

We can define nearly autoparallel maps  of la--spaces as an
anisotropic generalization (see also [24, 25]) of ng--[22]  and na--maps
[17--21]:
\begin{defin}
Nearly autoparallel maps, na--maps, of la--spaces are
defined as local 1--1
mappings of v--bundles, $\xi {\to} {\underline {\xi}},$ changing
every a--parallel on $\xi$ into a na--parallel on ${\underline {\xi}}.$
\end{defin}

Now we formulate the general conditions when
deformations (12) and (13) characterize na-maps :
Let a-parallel $l {\subset} U$~ is given by functions
$u^{\alpha}=u^{({\alpha})}({\eta}),   v^{\alpha}={{du^{\alpha}} \over d{\eta}}$,
${\eta}_{1} < {\eta} < {\eta}_{2}$,  satisfying equations (14). We suppose that to
this a--parallel corresponds a na--parallel ${\underline l} \subset
{\underline U}.$
given by the same parameterization in a common for a chosen na--map coordinate
system on $U$~ and ${\underline U}.$  This condition holds for vectors
${\underline v}^{\alpha}_{(1)}=v{\underline D}v^{\alpha}$~ and
$v^{\alpha}_{(2)}= v{\underline D}v^{\alpha}_{(1)}$ satisfying equality
\begin{equation}
 {\underline v}^{\alpha}_{(2)}= {\underline a}({\eta})v^{\alpha}+
{\underline b}({\eta}){\underline v}^{\alpha}_{(1)}   \label {f15}
\end{equation}
for some scalar
functions $ {\underline a}({\eta})$~ and ${\underline b}({\eta})$~
(see Definitions 4 and 5). Putting splittings (12) and (13) into expressions
for${\quad} {\underline v}^{\alpha}_{(1)}$~ and ${\underline v}^{\alpha}_{(2)}$~
 in (15) we obtain :
\widetext
\begin{equation}
 v^{\beta}v^{\gamma}v^{\delta}(D_{\beta}P^{\alpha}_{{.}{\gamma}{\delta}}+
P^{\alpha}_{{.}{\beta}{\tau}}P^{\tau}_{{.}{\gamma}{\delta}}+
Q^{\alpha}_{{.}{\beta}{\tau}}P^{\tau}_{{.}{\gamma}{\delta}})=
bv^{\gamma}v^{\delta}P^{\alpha}_{{.}{\gamma}{\delta}}+av^{\alpha},
 \label {f16}
\end{equation}
where
\begin{equation}
 b({\eta}, v)={\underline b}-3{\rho}, \qquad \mbox{and} \qquad
 a({\eta}, v)=
{\underline a}+{\underline b}{\rho}-v^{b}{\partial}_{b}{\rho}-{\rho}^{2}
  \label {f17}
\end{equation}
 are called the deformation parameters of na--maps.

The algebraic equations for the deformation of torsion
$Q^{\alpha}_{{.}{\beta}{\tau}}$
should be written as the compatibility conditions for a given nonmetricity
tensor
${\underline K}_{{\alpha}{\beta}{\gamma}}$~ on ${\underline {\xi}}$
( or as the metricity conditions if d--connection ${\underline D}_{\alpha}$~ on
${\underline {\xi}}$~ is required to be metric) :
\begin{eqnarray}
D_{\alpha}G_{{\beta}{\gamma}}-P^{\delta}_{{.}{\alpha}({\beta}}G_{{{\gamma})}
{\delta}}-{\underline K}_{{\alpha}{\beta}{\gamma}}=
Q^{\delta}_{{.}{\alpha}({\beta}}G_{{\gamma}){\delta}},  \label {f18}
\end{eqnarray}
where $({\quad})$ denotes the symmetric alternation.

So, we have proved this
\begin{theorem}
 The na--maps from la--space $\xi$ to la--space
$ {\underline {\xi}}$~ with a fixed common nonlinear connection
$ N^{a}_{j}(u)={\underline N}^{a}_{j}(u)$ and given d--connections,
${\Gamma}^{\alpha}_{{.}{\beta}{\gamma}}$~ on $\xi$~ and
 $ {\underline {\Gamma}}^{\alpha}_{{.}{\beta}{\gamma}}$~ on
$ {\underline {\xi}}$ are locally parametrized by the solutions of equations
(16) and (18) for every point $u^{\alpha}$~ and direction $v^{\alpha}$~ on
$U {\subset} {\xi}.$
\end{theorem}

We call (16) and (18) the basic equations for na--maps of la--spaces.
They generalize the corresponding Sinyukov's equations [22] for isotropic
spaces provided with symmetric affine connection structure.

\section { Classification of na--maps of la--spaces }

Na--maps are classed on possible polynomial parametrizations on variables
$v^{\alpha}$~ of deformations parameters $a$ and $b$ (see (16) and (17) ).
\begin{theorem}
 There are four classes of na--maps characterized
by corresponding deformation parameters and tensors and basic equations:
\begin{enumerate}
\item  for $na_{(0)}$--maps, ${\pi}_{(0)}$--maps,
$$P^{\alpha}_{{\beta}{\gamma}}(u)={\psi}
_{{(}{\beta}}{\delta}^{\alpha}_{{\gamma})}$$
(${\delta}^{\alpha}_{\beta}$~ is Kronecker symbol and ${\psi}_{\beta}=
{\psi}_{\beta}(u)$~ is a covariant vector d--field) ;
\item
 for $ na_{(1)}$--maps
$$ a(u,v)= a_{{\alpha}{\beta}}(u)v^{\alpha}v^{\beta},\quad
b(u,v)=b_{\alpha}(u)v^{\alpha}$$
and $P^{\alpha}_{{.}{\beta}{\gamma}}(u)$~ is the solution of equations
\widetext
\begin{eqnarray}
 D_{({\alpha}}P^{\delta}_{{.}{\beta}{\gamma})}+
P^{\tau}_{({\alpha}{\beta}}P^{\delta}_{{.}{\gamma}){\tau}}-
P^{\tau}_{({\alpha}{\beta}}Q^{\delta}_{{.}{\gamma}){\tau}} =
b_{({\alpha}}P^{{\delta}}_{{.}{\beta}{\gamma})}+a_{({\alpha}{\beta}}{\delta}
^{\delta}_{{\gamma})};  \label {f19}
\end{eqnarray}
\item  for $ na_{(2)}$--maps
$$a(u, v)= a_{\beta}(u) v^{\beta}, \quad b(u, v)= {{b_{{\alpha}{\beta}}
v^{\alpha}v^{\beta}}\over{{\sigma}_{\alpha}(u)v^{\alpha}}}, \quad
{\sigma}_{\alpha}v^{\alpha} {\neq}0,$$
$$
P^{\tau}_{{.}{\alpha}{\beta}}(u)={{\psi}_{({\alpha}}}{\delta}^{\tau}_{{\beta})}+
{\sigma}_{({\alpha}}F^{\tau}_{{\beta})}$$
and $F^{\alpha}_{\beta}(u)$~ is the solution of equations
\begin{eqnarray}
 {D}_{({\gamma}}F^{\alpha}_{{\beta})}+
F^{\alpha}_{\delta}F^{\delta}_{({\gamma}}{\sigma}_{{\beta})}-
Q^{\alpha}_{{.}{\tau}({\beta}}F^{\tau}_{{\gamma})} =
{\mu}_{({\beta}}F^{\alpha}_
{{\gamma})}+{\nu}_{({\beta}}{\delta}^{\alpha}_{{\gamma})}  \label {f20}
\end{eqnarray}
$ ({\mu}_{\beta}(u), {\nu}_{\beta}(u), {\psi}_{\alpha}(u),
{\sigma}_{\alpha}(u) $~ are covariant d--vectors) ;

\item for $na_{(3)}$--maps
$$
b(u,v)={{{\alpha}_{{\beta}{\gamma}{\delta}}v^{\beta}v^{\gamma}v^{\delta}}\over
{{\sigma}_{{\alpha}{\beta}}v^{\alpha}v^{\gamma}}},$$
$$
P^{\alpha}_{{.}{\beta}{\gamma}}(u)=
{\psi}_{({\beta}}{\delta}^{\alpha}_{{\gamma}
)}+
{\sigma}_{{\beta}{\gamma}}{\varphi}^{\alpha},$$
where ${\varphi}^{\alpha}$~ is the solution of equations
\begin{equation}
$$ D_{\beta}{\varphi}^{\alpha}={\nu}{\delta}^{\alpha}_{\beta}+
{\mu}_{\beta}{\varphi}^{\alpha}+
{\varphi}^{\gamma}Q^{\alpha}_{{.}{\gamma}{\delta}},   \label {f21}
\end{equation}
${\alpha}_{{\beta}{\gamma}{\delta}}(u), {\sigma}_{{\alpha}{\beta}}(u), {\psi}_
{\beta}(u), {\nu}(u)$~ and ${\mu}_{\beta}(u)$~ are d--tensors.
\end{enumerate}
\end{theorem}
The proof of the theorem is sketched in the Appendix A.

We point out that for ${\pi}_{(0)}$--maps we have not differential equations on
$P^{\alpha}_{{.}{\beta}{\gamma}}$ ( in the isotropic case one considers a
first order system of differential equations on metric [22]; we omit constructions
 with deformation of metric in this work).

To formulate invariant conditions for reciprocal na--maps (when  every
a-parallel on ${\underline {\xi}}$~ is also
transformed into na--parallel  on $\xi$ ) it is
convenient to introduce into consideration the curvature and Ricci tensors
defined for auxiliary connection ${\gamma}^{\alpha}_{{.}{\beta}{\gamma}}$~ :
$$r^{{.}{\delta}}
_{{\alpha}{.}{\beta}{\tau}}={\partial}_{[{\beta}}{\gamma}^{\delta}_
{{.}{\tau}]{\alpha}}+{\gamma}^{\delta}_{{.}{\rho}[{\beta}}{\gamma}^{\rho}_
{{.}{\tau}]{\alpha}} + {{\gamma}^{\delta}}_{\alpha \phi} {w^{\phi}}_{\beta \tau}$$
     and, respectively, $r_{{\alpha}{\tau}}=r^{{.}{\gamma}}
_{{\alpha}{.}{\gamma}{\tau}} $,
where $[\quad ]$ denotes antisymmetric alternation of indices, and
 to define values:
\widetext
$$^{(0)}T^{\mu}_{{.}{\alpha}{\beta}}=
{\Gamma}^{\mu}_{{.}{\alpha}{\beta}} - T^{\mu}_{{.}{\alpha}{\beta}}-
{1 \over (n+m + 1)}({\delta}^{\mu}_{({\alpha}}{\Gamma}^{\delta}_
{{.}{\beta}){\delta}}-{\delta}^{\mu}_{({\alpha}}T^{\delta}_
{{.}{\beta}){\gamma}}), $$
$${}^{(0)}{W}^{\cdot \tau}_{\alpha \cdot \beta \gamma} =
   {r}^{\cdot \tau}_{\alpha \cdot \beta \gamma} +
{1\over n+m+1} [ {{\gamma}}^{\tau}_{\cdot \varphi \tau}
 {\delta}^{\tau}_{( \alpha} {w^{\varphi}}_{\beta ) \gamma} -
( {\delta}^{\tau}_{\alpha}{r}_{[ \gamma \beta ]}  +
{\delta}^{\tau}_{\gamma} {r}_{[ \alpha \beta ]} -
{\delta}^{\tau}_{\beta} {r}_{[ \alpha \gamma ]} )] -$$
$$
{1\over {(n+m+1)}^2} [ {\delta}^{\tau}_{\alpha}
 ( 2 {{\gamma}}^{\tau}_{\cdot \varphi \tau}
 {w^{\varphi}}_{[ \gamma \beta ] } -
{{\gamma}}^{\tau}_{\cdot \tau [ \gamma }
{w^{\varphi}}_{\beta ] \varphi} ) + {\delta}^{\tau}_{\gamma}
 ( 2 {{\gamma}}^{\tau}_{\cdot \varphi \tau}
  {w^{\varphi}}_{\alpha \beta} -{{\gamma}}^{\tau}_{\cdot \alpha \tau}
{w^{\varphi}}_{\beta \varphi}) - $$
$${\delta}^{\tau}_{\beta}
 ( 2 {{\gamma}}^{\tau}_{\cdot \varphi \tau}
  {w^{\varphi}}_{\alpha \gamma} - {{\gamma}}^{\tau}_{\cdot \alpha \tau}
{w^{\varphi}}_{\gamma \varphi} ) ]$$

$${^{(3)}T}^{\delta}_{{.}{\alpha}{\beta}}=
{\gamma}^{\delta}_{{.}{\alpha}{\beta}}+
{\epsilon}{\varphi}^{\tau}{^{({\gamma})}D}_{\beta}q_{\tau}+
{1 \over n+m}({\delta}^{\gamma}_{\alpha}-
{\epsilon}{\varphi}^{\delta}q_{\alpha})[{\gamma}^{\tau}_{{.}{\beta}{\tau}}+
{\epsilon}{\varphi}^{\tau}{^{({\gamma})}D}_{\beta}q_{\tau}+$$
$${1 \over {n+m -1}}q_{\beta}({\epsilon}{\varphi}^{\tau}{\gamma}^{\lambda}_
{{.}{\tau}{\lambda}}+
{\varphi}^{\lambda}{\varphi}^{\tau}{^{({\gamma})}D}_{\tau}q_{\lambda})]-
{1 \over n+m}({\delta}^{\delta}_{\beta}-{\epsilon}{\varphi}^{\delta}
q_{\beta})[{\gamma}^{\tau}_{{.}{\alpha}{\tau}}+
{\epsilon}{\varphi}^{\tau}
{^{({\gamma})}D}_{\alpha}q_{\tau}+$$
$${1 \over {n+m -1}}q_{\alpha}({\epsilon}{\varphi}^{\tau}{\gamma}^{\lambda}_
{{.}{\tau}{\lambda}}+
{\varphi}^{\lambda}{\varphi}^{\tau}
{^{(\gamma)}D}
_{\tau}q_{\lambda})],$$

$$ {^{(3)}W}^{\alpha}{{.}{\beta}{\gamma}{\delta}}=
{\rho}^{{.}{\alpha}}_{{\beta}{.}{\gamma}{\delta}}+
{\epsilon}{\varphi}^{\alpha}q_{\tau}
{\rho}^{{.}{\tau}}_{{\beta}{.}{\gamma}{\delta}}
+({\delta}^{\alpha}_{\delta} -
{\epsilon}{\varphi}^{\alpha}q_{\delta})p_{{\beta}{\gamma}}-
({\delta}^{\alpha}_{\gamma} -
{\epsilon}{\varphi}^{\alpha}q_{\gamma})p_{{\beta}{\delta}}-
({\delta}^{\alpha}_{\beta}-{\epsilon}{\varphi}^{\alpha}q_{\beta})
p_{[{\gamma}{\delta}]},$$
$$(n+m-2)p_{{\alpha}{\beta}}=-{\rho}_{{\alpha}{\beta}} -
{\epsilon}q_{\tau}{\varphi}^{\gamma}
{\rho}^{{.}{\tau}}_{{\alpha}{.}{\beta}{\gamma}} +
{1 \over n+m}
[{\rho}_{{\tau}{.}{\beta}{\alpha}}^{{.}{\tau}} -
{\epsilon}q_{\tau}{\varphi}^{\gamma}
{\rho}^{{.}{\tau}}_{{\gamma}{.}{\beta}{\alpha}}+
{\epsilon}q_{\beta}{\varphi}^{\tau}{\rho}_{{\alpha}{\tau}}+$$
$$
{\epsilon}q_{\alpha}(-{\varphi}^{\gamma}{\rho}^{{.}{\tau}}
_{{\tau}{.}{\beta}{\gamma}}+
{\epsilon}q_{\tau}{\varphi}^{\gamma}{\varphi}^{\delta}{\rho}^{{.}{\tau}}
_{{\gamma}{.}{\beta}{\delta}}]), $$
where $q_{\alpha}{\varphi}^{\alpha}={\epsilon}=\pm 1 ,$
$$
{{\rho}^{\alpha}}_{\beta \gamma \delta} =
r^{\cdot \alpha}_{\beta \cdot \gamma \delta}  +
{1\over 2}( {\psi}_{(\beta} {\delta}^{\alpha}_{\varphi )} +
 {\sigma}_{\beta \varphi}{\varphi}^{\tau} ) {w^{\varphi}}_{\gamma \delta}$$
 ( for a similar value on $\underline {\xi}$ we write ${\quad}
{\underline {\rho}}^{\alpha}_{\cdot \beta \gamma \delta} =
{\underline r}^{\cdot \alpha}_{\beta \cdot \gamma \delta} -
{1\over 2}( {\psi}_{(\beta} {\delta}^{\alpha}_{{\varphi})} -
 {\sigma}_{\beta \varphi}{\varphi}^{\tau} )
{w^{\varphi}}_{\gamma \delta}{\quad})$ and ${\rho}_{\alpha \beta} =
{\rho}^{\tau}_{\cdot \alpha \beta \tau} .$


Similar values, $$^{(0)}{\underline T}^{\alpha}_{{.}{\beta}{\gamma}},
^{(0)}{\underline W}^{\nu}_{{.}{\alpha}{\beta}{\gamma}}, {\hat T}^{\alpha}
_{{.}{\beta}{\gamma}},
{\check T}^{\alpha}_
{{.}{\beta}{\tau}},
{\hat W}^{\delta}_{{.}{\alpha}{\beta}{\gamma}}, {\check W}^{\delta}_
{{.}{\alpha}{\beta}{\gamma}}, ^{(3)}{\underline T}^{\delta}
_{{.}{\alpha}{\beta}},$$ and $^{(3)}{\underline W}^
{\alpha}_{{.}{\beta}{\gamma}{\delta}}  $ are given, correspondingly,
by auxiliary connections ${\quad}{\underline {\Gamma}}^{\mu}_{{.}{\alpha}{\beta}},$~
$$ {\star {\gamma}}^{\alpha}_{{.}{\beta}{\lambda}}={\gamma}^{\alpha}
_{{.}{\beta}{\lambda}} + {\epsilon}F^{\alpha}_{\tau}{^{({\gamma})}D}_{({\beta}}
F^{\tau}_{{\lambda})}, \quad
{\check {\gamma}}^{\alpha}_{{.}{\beta}{\lambda}}=
{\widetilde {\gamma}}^{\alpha}_{{.}{\beta}{\lambda}} + {\epsilon}F^{\lambda}
_{\tau}
{\widetilde D}_{({\beta}}F^{\tau}_{{\lambda})},$$
$${\widetilde {\gamma}}^{\alpha}_{{.}{\beta}{\tau}}={\gamma}^{\alpha}
_{{.}{\beta}{\tau}}+ {\sigma}_{({\beta}}F^{\alpha}_{{\tau})}, \quad
{\hat {\gamma}}^{\alpha}_{{.}{\beta}{\lambda}}={\star {\gamma}}^{\alpha}_
{{.}{\beta}{\lambda}} + {\widetilde {\sigma}}_{({\beta}}{\delta}^{\alpha}_
{{\lambda})},$$
where ${\widetilde {\sigma}}_{\beta}={\sigma}_{\alpha}F^{\alpha}_{\beta}.$
\begin{theorem}
 Four classes of reciprocal na--maps of la--spaces
are characterized by corresponding invariant criterions:
\begin{enumerate}
\item for a--maps $^{(0)}T^{\mu}_{{.}{\alpha}{\beta}}=
              ^{(0)}{\underline T}^{\mu}_{{.}{\alpha}{\beta}}, $
\begin{equation}
{}^{(0)}W^{\delta}_{{.}{\alpha}{\beta}{\gamma}}= ^{(0)}{\underline W}
^{\delta}_{{.}{\alpha}{\beta}{\gamma}}; \label {f22}
\end{equation}
\item for $na_{(1)}$--maps
\begin{equation}
3({^{({\gamma})}D}_{\lambda}P^{\delta}_{{.}{\alpha}{\beta}}+
P^{\delta}_{{.}{\tau}{\lambda}}P^{\tau}_{{.}{\alpha}{\beta}})=
r^{{.}{\delta}}_{({\alpha}{.}{\beta}){\lambda}} -
{\underline r}^{{.}{\delta}}_{({\alpha}{.}{\beta}){\lambda}}+ \label {f23}
\end{equation}
$$[T^{\delta}_{{.}{\tau}({\alpha}}
P^{\tau}_{{.}{\beta}{\lambda})}+
Q^{\delta}_{{.}{\tau}({\alpha}}
P^{\tau}_{{.}{\beta}{\lambda})}+
b_{({\alpha}}P^{\delta}_{{.}{\beta}{\lambda})}+
{\delta}^{\delta}_{({\alpha}}a_{{\beta}{\lambda})}];  $$

\item for $na_{(2)}$--maps ${\hat T}^{\alpha}_{{.}{\beta}{\tau}}=
                      {\star T}^{\alpha}_{{.}{\beta}{\tau}}, $
\begin{equation}
 {\hat W}^{\delta}_{{.}{\alpha}{\beta}{\gamma}}=
 {\star W}^{\delta}_{{.}{\alpha}{\beta}{\gamma}}; \label{f24}
\end{equation}
\item for $na_{(3)}$--maps $^{(3)}T^{\alpha}_{{.}{\beta}{\gamma}}=
          ^{(3)}{\underline T}^{\alpha}_{{.}{\beta}{\gamma}},$
\begin{equation}
{}^{(3)}W^{\alpha}_{{.}{\beta}{\gamma}{\delta}}= ^{(3)}
{\underline W}^{\alpha}_{{.}{\beta}{\gamma}{\delta}}.  \label {f25}
\end{equation}
\end{enumerate}
\end{theorem}
Proof of this  theorem is sketched in the Appendix B.

For the particular case of $na_{(3)}$--maps when ${\psi}_{\alpha}=0 ,
 {\varphi}_{\alpha} = g_{\alpha \beta} {\varphi}^{\beta} =
 {\delta \over \delta u^{\alpha}} ( \ln {\Omega} ) , {\Omega}(u) > 0 $
and ${\sigma}_{\alpha \beta} = g_{\alpha \beta}$ we define a subclass
of conformal transforms ${\underline g}_{\alpha \beta} (u) = {\Omega}^2 (u)
 g_{\alpha \beta}$ which, in consequence of the fact that d--vector
 ${\varphi}_{\alpha}$ must satisfy equations
(21), generalizes the class of concircular transforms (see [22] for references
and details on concircular mappings of Riemannaian spaces) .

We emphasize that basic na--equations (19)--(21) are systems of first order
partial differential equations. The study  of their geometrical properties and
definition of integral varieties, general and particular solutions are
possible by using the formalism of Pffaf systems [17]. Here we point out
that by using algebraic methods we can always verify if systems of
na--equations  of type (19)--(21) are, or not, involute, even to find their
explicit solutions it is a difficult task (see more detailed considerations
for isotropic ng--maps in [22] and, on language of Pffaf systems for na--maps,
in [20]). We can also formulate the Cauchy problem for na--equations
on $\xi$~ and choose deformation parameters (17) as to make involute
mentioned equations for the case of maps to a given background space
${\underline {\xi}}$. If a solution, for example, of
$na_{(1)}$--map equations exists, we say that space $\xi$ is $ na_{(1)}$--projective
to space ${\underline {\xi}}.$  In general, we have to introduce chains of
na--maps in order to obtain involute systems of equations for maps
(superpositions of na-maps) from $\xi$ to ${\underline {\xi}}:$
$$U \buildrel {ng<i_{1}>} \over \longrightarrow  {U_{\underline 1}}
\buildrel ng<i_2> \over \longrightarrow \cdots
\buildrel ng<i_{k-1}> \over \longrightarrow U_{\underline {k-1}}
\buildrel ng<i_k> \over \longrightarrow {\underline U} $$
where $U \subset {\xi}, U_{\underline 1} \subset {\xi}_{\underline 1}, \ldots ,
U_{k-1} \subset {\xi}_{k-1}, {\underline U} \subset {\xi}_k$
with corresponding splittings of auxiliary symmetric connections
$${\underline {\gamma}}^{\alpha}_{.{\beta}{\gamma}}= _{<i_1>}P^{\alpha}_{.{\beta}{\gamma}}+
_{<i_2>}P^{\alpha}_{.{\beta}{\gamma}}+ \cdots +
 _{<i_k>}P^{\alpha}_{.{\beta}{\gamma}}$$
and torsion
$${\underline T}^{\alpha}_{.{\beta}{\gamma}}=
T^{\alpha}_{.{\beta}{\gamma}}+ _{<i_1>}Q^{\alpha}_{.{\beta}{\gamma}}+
_{<i_2>}Q^{\alpha}_{.{\beta}{\gamma}}+ \cdots
+ _{<i_k>}Q^{\alpha}_{.{\beta}{\gamma}}$$
where cumulative indices $<i_1>=0, 1, 2, 3, $ denote possible types of na--maps.
\begin{defin}
 Space $\xi$~ is nearly conformally projective to
space ${\underline {\xi}}, {\quad} nc : {\xi} {\to} {\underline {\xi}},$~ if
there is
a finite chain of na--maps from $\xi$~ to ${\underline {\xi}}.$
\end{defin}

For nearly conformal maps we formulate :
\begin{theorem}
 For every fixed triples $(N^{a}_{j}, {\Gamma}^{\alpha}_{{.}{\beta}{\gamma}},
 U \subset {\xi} $ and $(N^{a}_{j}, {\underline {\Gamma}}^
{\alpha}_{{.}{\beta}{\gamma}}$, ${\underline U} \subset {\underline {\xi}})$,
components of nonlinear connection, d--connection and d--metric
being of class $C^{r}(U), C^{r}({\underline U})$,
  $r>3,$ there is a finite chain
of na--maps $nc : U \to {\underline U}.$
\end{theorem}
Proof is similar to that for isotropic maps [18,20,24]. For analytic functions
it is a direct consequence from the Cauchy-- Kowalewski theorem [31].

Now we introduce the concept of the Category of la--spaces,
${\cal C}({\xi}).$  The elements
of ${\cal C}({\xi})$ consist from $Ob{\cal C}({\xi})=\{{\xi}, {\xi}_{<i_{1}>},
 {\xi}_{<i_{2}>},{\ldots}, \}$  being la--spaces, for
simplicity in this work, having common N--connection structures,
and $Mor {\cal C}({\xi})=\{ nc ({\xi}_{<i_{1}>}, {\xi}_{<i_{2}>})\}$
being chains of na--maps interrelating la--spaces. We point out
that we can consider equivalent models of physical theories on every
object of ${\cal C}({\xi})$  (see details for isotropic gravitational
models in [17--21] and anisotropic gravity in [24,25] ). One of the main purposes
of this paper is to develop a d--tensor and variational formalism on
${\cal C}({\xi}),$ i.e. on la--multispaces, interrelated with
nc--maps. Taking into account the distinguished character of geometrical
objects    on la--spaces we call tensors on ${\cal C}({\xi})$ as
distinguished tensors on la--space Category, or dc--tensors.

Finally, we emphasize that presented in that section definitions and theorems
can be generalized for  v--bundles  with arbitrary given structures
of nonlinear connection, linear d--connection and metric structures. Proofs
are similar to those presented in [21,22] but rather cumbersome.

\section{ Nearly Autoparallel Tensor--Integral on La--spaces}

The aim of this section is to define tensor integration not only for
bitensors, objects defined on the same curved space, but for dc--tensors,
defined on two spaces, $\xi$ and ${\underline {\xi}}$, even it is necessary
on la--multispaces. A. Mo\'or tensor--integral formalism [14] having a lot
of applications in classical and quantum gravity [32-34,15] was
extended for locally isotropic multispaces in [17,18]. The unispacial
locally anisotropic version is given in [28,29].

 Let $T_{u}{\xi}$~ and
$T_{\underline u}{\underline {\xi}}$ be tangent spaces in corresponding
points $u {\in} U {\subset} {\xi}$ and ${\underline u} {\in} {\underline U}
{\subset} {\underline {\xi}}$ and, respectively, $T^{\ast}_{u}{\xi}$ and
$T^{\ast}_{\underline u}{\underline {\xi}} $ be their duals (in general,
in this section we shall not consider that a common coordinatization is
introduced for open regions $U$ and ${\underline U}$ ). We call as the dc--tensors
on the pair of spaces (${\xi}, {\underline {\xi}}$ ) the elements of
distinguished  tensor algebra
$$ ( {\otimes}_{\alpha} T_{u}{\xi}) {\otimes} ({\otimes}_{\beta} T^{\ast}_{u}
{\xi}) {\otimes}({\otimes}_{\gamma} T_{\underline u}{\underline {\xi}})
{\otimes} ({\otimes}_{\delta} T^{\ast}_{\underline u} {\underline {\xi}})$$
defined over the space ${\xi}{\otimes} {\underline {\xi}},  $ for a given
$nc : {\xi} {\to} {\underline {\xi}} $.

We admit the convention that underlined and non--underlined indices refer,
 respectively, to the points ${\underline u}$ and $u$. Thus $Q_{{.}{\underline
{\alpha}}}^{\beta}, $  for instance, are the components of dc--tensor
$Q{\in} T_{u}{\xi} {\otimes} T_{\underline u}{\underline {\xi}}.$

Now, we define the transport dc--tensors. Let open regions $U$ and ${\underline U}$
be homeomorphic to sphere ${\cal R}^{2n}$ and introduce isomorphism
${\mu}_{{u},{\underline u}}$  between
$T_{u}{\xi}$ and $T_{\underline u}{\underline {\xi}}$ (given by map
$nc : U {\to} {\underline U}).$ We consider that for every d--vector
$ v^{\alpha} {\in} T_{u}{\xi}$ corresponds the vector
 ${\mu}_{{u},{\underline u}}
(v^{\alpha})=v^{\underline {\alpha}} {\in} T_{\underline u}{\underline {\xi}},$
with components $v^{\underline {\alpha}}$ being linear functions of
 $v^{\alpha}$:
$$ v^{\underline {\alpha}}=h^{\underline {\alpha}}_{\alpha}(u, {\underline u})
v^{\alpha}, \quad v_{\underline {\alpha}}=
h^{\alpha}_{\underline {\alpha}}({\underline u}, u)v_{\alpha},$$
where $h^{\alpha}_{\underline {\alpha}}({\underline u}, u)$  are the components
of dc--tensor associated with ${\mu}^{-1}_{u,{\underline u}}$. In a similar
manner we have
$$ v^{\alpha}=h^{\alpha}_{\underline {\alpha}}({\underline u}, u)
v^{\underline {\alpha}}, \quad v_{\alpha}=h^{\underline {\alpha}}_{\alpha}
(u, {\underline u})v_{\underline {\alpha}}.$$

In order to reconcile just presented definitions and to assure the identity
for trivial maps ${\xi} {\to} {\xi}, u={\underline u},$  the transport
dc-tensors must satisfy conditions :
$$ h^{\underline {\alpha}}_{\alpha}(u, {\underline u})h^{\beta}
_{\underline {\alpha}}({\underline u}, u)={\delta}^{\beta}_{\alpha},
h^{\underline {\alpha}}_{\alpha}(u,{\underline u})
h^{\alpha}_{\underline {\beta}}({\underline u}, u)={\delta}^
{\underline {\alpha}}_{\underline {\beta}}$$
and ${\lim}_{{({\underline u}{\to}u})}h^{\underline {\alpha}}
_{\alpha}(u,{\underline u}) ={\delta}_{\alpha}^{\underline {\alpha}}, \quad
     {\lim}_{{({\underline u}{\to}u})}h^{\alpha}_{\underline {\alpha}}({\underline u},u)=
{\delta}^{\alpha}_{\underline {\alpha}}.$

Let $ {\overline S}_{p} {\subset} {\overline U} {\subset} {\overline {\xi}}$
is a homeomorphic to $p$-dimensional sphere and suggest that chains of na--maps
are used to connect  regions :
$$ U \buildrel nc_{(1)} \over \longrightarrow {\overline S}_p
     \buildrel nc_{(2)} \over \longrightarrow {\underline U}.$$

\begin{defin}
 The tensor integral in ${\overline u} {\in }
{\overline S}_{p}$ of a dc--tensor $N^{{.}{\gamma}{.}{\underline {\kappa}}}
_{{\varphi}{.}{\underline {\tau}}{.}{\overline {\alpha}}_
{1}{\cdots}{\overline {\alpha}}_{p}}$   $({\overline u}, u),$
completely antisymmetric on the indices ${{\overline {\alpha}}_{1}},{\ldots},
{\overline {\alpha}}_{p},$  over domain ${\overline S}_{p},$  is defined as
$$ N^{{.}{\gamma}{.}{\underline {\kappa}}}_{{\varphi}{.}{\underline {\tau}}}
({\underline u}, u)=I^{\underline U}_{({\overline S}
_{p})}N^{{.}{\gamma}{.}{\overline {\kappa}}}
_{{\varphi}{.}{\overline {\tau}}{.}{\overline {\alpha}}
_{1}{\ldots}{\overline {\alpha}}_{p}}({\overline u}, {\underline u})
dS^{{\overline {\alpha}}_{1}{\ldots}{\overline {\alpha}}_{p}}=$$
\begin{equation}
{\int}_{({\overline S}_{p})}h^{\overline {\tau}}_{\underline {\tau}}
({\underline u}, {\overline u})h^{\underline {\kappa}}_{\overline {\kappa}}
({\overline u}, {\underline u})N^{{.}{\gamma}{.}{\overline {\kappa}}}_
{{\varphi}{.}{\overline {\tau}}{.}{\overline {\alpha}}
_{1}{\cdots}{\overline {\alpha}}_{p}}
({\overline u}, u)d{\overline S}^{{\overline {\alpha}}
_{1}{\cdots}{\overline {\alpha}}_{p}},                          \label {f26}
\end{equation}
where $dS^{{\overline {\alpha}}_{1}{\cdots}{\overline {\alpha}}_{p}}=
{\delta}u^{{\overline {\alpha}}_{1}}{\land}{\cdots}{\land}{\delta}u^
{\overline {\alpha}}_{p}$.
\end{defin}

Let suppose that transport dc--tensors  $h^{\underline {\alpha}}_{\alpha}$~
and $h^{\alpha}_{\underline {\alpha}}$~ admit covariant derivations of order
two and postulate  existence of deformation dc--tensor $B^{{..}{\gamma}}
_{{\alpha}{\beta}}(u, {\underline u})$~ satisfying relations
\begin{equation}
D_{\alpha}h^{\underline {\beta}}_{\beta}(u, {\underline u})=
B^{{..}{\gamma}}_{{\alpha}{\beta}}(u, {\underline u})
h^{\underline {\beta}}_{\gamma}(u, {\underline u})  \label {f27}
\end{equation}
and, taking into account that $D_{\alpha}{\delta}^{\beta}_{\gamma}=0, $

$$ D_{\alpha}h^{\beta}_{\underline {\beta}}({\underline u}, u)=
-B^{{..}{\beta}}_{{\alpha}{\gamma}}(u, {\underline u})
h^{\gamma}_{\underline {\beta}}({\underline u}, u).$$
 By using formulas (4)  and (5) for torsion and, respectively,
curvature of connection ${\Gamma}^{\alpha}_{{\beta}{\gamma}}$~
we can calculate next commutators:
\begin{equation}
 D_{[{\alpha}}D_{{\beta}]}h^{\underline {\gamma}}_{\gamma}=-
(R^{{.}{\lambda}}_{{\gamma}{.}{\alpha}{\beta}}+T^{\tau}_{{.}{\alpha}{\beta}}
B^{{..}{\lambda}}_{{\tau}{\gamma}})h^{\underline {\gamma}}_{\lambda}.\label {f28}
\end{equation}
On the other hand from (27) one follows that
\begin{equation}
 D_{[{\alpha}}D_{{\beta}]}h^{\underline {\gamma}}_{\gamma}=(
D_{[{\alpha}}B_{{\beta}]{\gamma}}^{{..}{\lambda}}+
B^{{..}{\lambda}}_{[{\alpha}{\vert}{\tau}{\vert}{.}}B
^{{..}{\tau}}_{{\beta}]{\gamma}{.}})h^{\underline {\gamma}}_{\lambda},\label{f29}
\end{equation}
where ${\vert}{\tau}{\vert}$~ denotes that index ${\tau}$~ is excluded
from the action
of antisymmetrization $[{\quad}]$. From (28) and (29) we obtain
\begin{equation}
 D_{[{\alpha}}B_{{\beta}]{\gamma}{.}}^{{..}{\lambda}}+
B_{[{\beta}{\vert}{\gamma}{\vert}}B^{{..}{\lambda}}_{{\alpha}]{\tau}}=
(R^{{.}{\lambda}}_{{\gamma}{.}{\alpha}{\beta}}+T^{\tau}_
{{.}{\alpha}{\beta}}
B^{{..}{\lambda}}_{{\tau}{\gamma}}).  \label {f30}
\end{equation}

Let ${\overline S}_{p}$~ be the boundary of ${\overline S}_{p-1}$.
 The Stoke's
type formula for tensor--integral (26) is defined as
\widetext
\begin{equation}
$$ I_{{\overline S}_{p}}N_{{\varphi}{.}{\overline {\tau}}{.}
{\overline {\alpha}}_{1}
{\ldots}{\overline {\alpha}}_{p}}^{{.}{\gamma}{.}{\overline {\kappa}}}
dS^{{\overline {\alpha}}_{1}
{\ldots}{\overline {\alpha}}_{p}}=$$
I_{{\overline S}_{p+1}}{^{{\star}{(p)}}{\overline D}}_{[{\overline {\gamma}}
{\vert}}
N^{{.}{\gamma}{.}{\overline {\kappa}}}
_{{\varphi}{.}{\overline {\tau}}{.}{\vert}{\overline {\alpha}}_{1}{\ldots}
{{\overline {\alpha}}_{p}]}}dS^{{\overline {\gamma}}{\overline {\alpha}}_{1}{\ldots}
{\overline {\alpha}}_{p}}, \label {f31}
\end{equation}
where
$$ {^{{\star}{(p)}}D}_{[{\overline {\gamma}}{\vert}}N^
{{.}{\gamma}{.}{\overline {\kappa}}}_
{{\varphi}{.}{\overline {\tau}}{.}{\vert}{\overline {\alpha}}_{1}{\ldots}
{\overline {\alpha}}_{p}]}=$$
\begin{equation}
D_{[{\overline {\gamma}}{\vert}}N^{{.}{\gamma}{.}{\overline {\kappa}}}
_{{\varphi}{.}{\overline {\tau}}{.}{\vert}{\overline {\alpha}}_{1}{\ldots}
{\overline {\alpha}}_{p}]}+
pT^{\underline {\epsilon}}_{{.}[{\overline {\gamma}}{\overline {\alpha}}_{1}{\vert}}
N^{{.}{\gamma}{.}{\overline {\kappa}}}_{{\varphi}{.}{\overline {\tau}}{.}
{\overline {\epsilon}}{\vert}{\overline {\alpha}}_{2}{\ldots}
{\overline {\alpha}}_{p}]} -
B_{[{\overline {\gamma}}{\vert}{\overline {\tau}}}^
{{..}{\overline {\epsilon}}}
N^{{.}{\gamma}{.}{\overline {\kappa}}}_{{\varphi}{.}{\overline {\epsilon}}{.}
{\vert}{\overline {\alpha}}_{1}{\ldots}{\overline {\alpha}}_{p}]}+
B_{[{\overline {\gamma}}{\vert}{\overline {\epsilon}}}^{..{\overline {\kappa}}}
N^{{.}{\gamma}{.}{\overline {\epsilon}}}_{{\varphi}{.}{\overline {\tau}}{.}
{\vert}{\overline {\alpha}}_{1}{\ldots}{\overline {\alpha}}_{p}]}.
 \label {f32} \end{equation}

 We define the dual element of
the hypersurfaces element $ dS^{{j}_1{\ldots}{j}_p}$ as
\begin{equation}
 d{\cal S}_{{\beta}_{1}{\ldots}{\beta}_{q-p}}=
{1 \over {p!}}{\epsilon}_{{\beta}_{1}{\ldots}{\beta}_{k-p}{\alpha}_{1}{\ldots}
{\alpha}_{p}}dS^{{\alpha}_{1}{\ldots}{\alpha}_{p}}, \label {f33}
\end{equation}

where ${\epsilon}_{{\gamma}_{1}{\ldots}{\gamma}_{q}}$  is completely
antisymmetric on its indices and
$${\epsilon}_{1 2 {\ldots} (n+m)}=\sqrt{{\vert}G{\vert}},
 G=det {\vert}G_{{\alpha}{\beta}{\vert}}, $$
$G_{{\alpha}{\beta}} $  is taken from (7). The dual of dc--tensor
$N^{{.}{\gamma}{\overline {\kappa}}}_{{\varphi}{.}{\overline {\tau}}{.}
{\overline {\alpha}}_{1}{\ldots}{\overline {\alpha}}_{p}}$
is defined as the dc--tensor
${\cal N}^{{.}{\gamma}{.}{\overline {\kappa}}{\overline {\beta}}_{1}{\ldots}
{\overline {\beta}}_{n+m-p}}_{{\varphi}{.}{\overline {\tau}}}$
satisfying
\begin{equation}
 N^{{.}{\gamma}{.}{\overline {\kappa}}}_{{\varphi}{.}{\overline {\tau}}{.}
{\overline {\alpha}}_{1}{\ldots}{\overline {\alpha}}_{p}}=
{1 \over {p!}}{\cal N}_{{\varphi}{.}{\overline {\tau}}}^
{{.}{\gamma}{.}{\overline {\kappa}}{\overline {\beta}}_
{1}{\ldots}{\overline {\beta}}_{n+m-p}}
{\epsilon}_{{\overline {\beta}}_{1}{\ldots}{\overline {\beta}}_{n+m-p}
{\overline {\alpha}}_{1}{\ldots}{\overline {\alpha}}_{p}}.  \label {f34}
\end{equation}
Using (26), (33) and (34) we can write
\begin{equation}
 I_{{\overline S}_{p}}N^{{.}{\gamma}{.}{\overline {\kappa}}}
_{{\varphi}{.}{\overline {\tau}}{.}{\overline {\alpha}}_{1}{\ldots}
{\overline {\alpha}}_{p}}dS^{{\overline {\alpha}}
_{1}{\ldots}{\overline {\alpha}}_{p}}=
{\int}_{{\overline S}_{p+1}}{^{\overline p}D}_{\overline {\gamma}}{\cal N}
_{{\varphi}{.}{\overline {\tau}}}^
{{.}{\gamma}{.}{\overline {\kappa}}{\overline {\beta}}_{1}
{\ldots}{\overline {\beta}}_{n+m-p-1}{\overline {\gamma}}}
d{\cal S}_{{\overline {\beta}}_{1}{\ldots}{\overline {\beta}}_{n+m-p-1}},
 \label {f35}
\end{equation}
where
 $$ {^{\overline p}D}_{\overline {\gamma}}{\cal N}_
{{\varphi}{.}{\overline {\tau}}}
^{{.}{\gamma}{.}{\overline {\kappa}}{\overline {\beta}}_
{1}{\ldots}{\overline {\beta}}
_{n+m-p-1}{\overline {\gamma}}}=$$
$$
{\overline D}_{\overline {\gamma}}{\cal N}_{{\varphi}{.}{\overline {\tau}}}^
{{.}{\gamma}{.}{\overline {\kappa}}{\overline {\beta}}_
{1}{\ldots}{\overline {\beta}}_{n+m-p-1}
{\overline {\gamma}}}+
(-1)^{(n+m-p)}(n+m-p+1)T^{[{\overline {\epsilon}}}_
{{.}{\overline {\gamma}}{\overline {\epsilon}}}
{\cal N}_{{\varphi}{.}{{\overline \tau}}}^
{{.}{\vert}{\gamma}{.}{\overline {\kappa}}{\vert}
{\overline {\beta}}_{1}{\ldots}{\overline {\beta}}_{n+m-p-1}]
{\overline {\gamma}}}-$$
$$B_{{\overline {\gamma}}{\overline {\tau}}}^{{..}{\overline {\epsilon}}}
{\cal N}_{{\varphi}{.}{\overline {\epsilon}}}^
{{.}{\gamma}{.}{\overline {\kappa}}
{\overline {\beta}}_{1}{\ldots}{\overline {\beta}}_{n+m-p-1}{\overline {\gamma}}}+
B_{{\overline {\gamma}}{\overline {\epsilon}}}^{{..}{\overline {\kappa}}}
{\cal N}_{{\varphi}{.}{\overline {\tau}}}^{{.}{\gamma}{.}{\overline {\epsilon}}
{\overline {\beta}}_{1}{\ldots}{\overline {\beta}}_
{n+m-p-1}{\overline {\gamma}}}.$$
To verify the equivalence of (34) and (35) we must take in consideration that
$$D_{\gamma}{\epsilon}_{{\alpha}_{1}{\ldots}{\alpha}_{k}}=0 \qquad \mbox{and} \qquad
{\epsilon}_{{\beta}_{1}{\ldots}{\beta}_{n+m-p}{\alpha}_{1}{\ldots}{\alpha}_{p}}
{\epsilon}^{{\beta}_{1}{\ldots}{\beta}_{n+m-p}{\gamma}_{1}{\ldots}{\gamma}_{p}}=
p!(n+m-p)!{\delta}^{[{\gamma}_{1}}_{{\alpha}_{1}}{\cdots}{\delta}^
{{\gamma}_{p}]}
_{{\alpha}_{p}}.$$
The developed in this section tensor integration formalism will be used in the
next section for definition of conservation laws on spaces with local anisotropy.

\section{ Conservation Laws on Locally Anisotropic Spaces}

To define conservation laws on locally anisotropic spaces is a challenging task
because of absence of global and local groups of automorphisms of such spaces.
Our main idea is to use chains of na--maps from a given, called hereafter as the
fundamental la--space to an auxiliary one with trivial curvatures and torsions
admitting a global group of automorphisms.  The aim of this section is to present
a brief introduction into the la--gravity (see [1,2] as basic references and
[11,12] for further developments for gauge like and spinor la--gravity) and
 formulate classes of conservation laws by using dc--objects and tensor--integral
values, na--maps and variational calculus on the Category of la--spaces.

\subsection{Locally Anisotropic Gravity in Vector Bundles}

 The Einstein equations on a v--bundle $\xi,$
associated to a metric d--connection $D$, with locally adapted coefficients
\widetext
$${{\Gamma}^{\alpha}}_{\beta \gamma} =$$
\begin{equation}
 ( {{\Gamma}^i}_{jk} = {L^i}_{jk} ,
{{\Gamma}^i}_{ja} = {C^i}_{ja} , {{\Gamma}^i}_{aj} = 0 , {{\Gamma}^i}_{ab} = 0,
 {{\Gamma}^a}_{jk} = 0, {{\Gamma}^a}_{jb} = 0, {{\Gamma}^a}_{bk} = {L^a}_{bk} ,
{{\Gamma}^a}_{bc} = {C^a}_{bc} ), \label {f36}
\end{equation}
where coefficients
\begin{equation}
{L^i}_{jk} =
 {1 \over 2} g^{ih} ({\delta g_{jh} \over \delta x^k} +
{\delta g_{kh} \over \delta x^j} - {\delta g_{jk} \over \delta x^h} +
g_{kl} {T^l}_{hj} + g_{lj} {T^l}_{hk} - g_{hi} {T^l}_{jk} ), \label {f37}
\end{equation}
$$ {L^a}_{bi} = {\partial N^a_i \over y^b} + {1\over 2} h^{ac}
 ({\delta h_{bc}\over \delta x^i} - {\partial N^d_i \over \partial y^b} h_{dc}
- {\partial N^d_i \over \partial y^c} h_{db} ) ,$$
$$ {C^i}_{jc} = {1\over 2} g^{ik} {\partial g_{jk} \over \partial y^c},$$
$$ {C^a}_{bc} =
{1\over 2} h^{ad} ( {\partial h_{db} \over \partial y^c} +
{\partial h_{dc} \over \partial y^b} - {\partial h_{bc} \over \partial y^d} +
h_{be} {S^e}_{bc} + h_{ce} {S^e}_{db} - h_{de} {S^e}_{bc})$$
are constructed with respect to locally adapted coefficients of metric (8),
 N--connection (1),(2) and presribed torsions ${T^i}_{jk}$ and ${S^a}_{bc},$
which in turn, hand in hand with (36), define the torsion (see formulas (4))
\begin{equation}
{{\bf T}^{\alpha}}_{\beta \gamma} = {{\Gamma}^{\alpha}}_{\beta \gamma} -
{{\Gamma}^{\alpha}}_{\gamma \beta} + {{w^{\alpha}}_{\beta \gamma}} \label {f38}
\end{equation}
having following components
$${{\bf T}^i}_{jk} = {T^i}_{jk}, {{\bf T}^i}_{ja} = {C^i}_{ja} ,
 {{\bf T}^i}_{aj} = - {C^i}_{ja}, {{\bf T}^i}_{ja} = 0 ,
{{\bf T}^a}_{bc} = {S^a}_{bc},$$
$${{\bf T}^a}_{ij} = {\delta N^a_i \over \delta x^j} - {\delta N^a_j \over
\delta x^i} ,
 {{\bf T}^a}_{bi} = {P^a}_{bi} =
 {\partial N^a_j \over \partial y^b} -
{{L^a}_{bj}} , {{{\bf T}^a}_{ib}} = - {{P^a}_{bi}} , $$
written by Miron and Anastasiei [1,2] for a generalization of general
relativity to the case of la--spaces. They introduced the Ricci tensor
${{\bf R}_{\beta \gamma}} = {\bf R}^{\cdot \alpha}_{\beta \cdot \gamma \alpha}$
constructed in a usual manner by using curvature of connection (36) (see
similar formulas (5)) having nonvanishing components:
\begin{eqnarray}
{{{R_h}^i}_{jk}} =
{\delta {L^i}_{hj} \over \delta x^k} -
{\delta {L^i}_{hk} \over \delta x^j} + {L^m}_{hj} {L^i}_{mk} - {L^m}_{hk}
{L^i}_{mj} + {C^i}_{ha} {R^a}_{jk} , \label {f39}
\end{eqnarray}
$${{{R_b}^a}_{jk}} =
 {\delta {L^a}_{bj} \over \delta x^k} -
{\delta {L^a}_{bk} \over \delta x^j} + {L^c}_{bj} {L^a}_{ck} - {L^c}_{bk}
{L^a}_{cj} + {C^a}_{bc} {R^c}_{jk} ,$$
$${{{P_j}^i}_{ka}} =
 {\partial {L^i}_{jk} \over \partial y^a} -
({\partial {C^i}_{ja}\over \partial x^k} + {L^i}_{lk}{C^l}_{ja} -
{L^l}_{jk} {C^i}_{la} - {L^c}_{ak} {{C^i}_{jc}}) + {C^i}_{jb} {P^b}_{ka} ,$$
$${{{P_b}^c}_{ka}} =
{\partial {L^c}_{bk} \over \partial y^a} -
({\partial {C^c}_{ba}\over \partial x^k} + {L^c}_{dk}{C^d}_{ba} -
{L^d}_{bk} {C^c}_{da} - {L^d}_{ak} {{C^c}_{bd}}) + {C^c}_{bd} {P^d}_{ks} ,$$
$${{S_j}^i}_{bc} =
 {\partial {C^i}_{jb} \over \partial y^c} -
{\partial {C^i}_{jc} \over \partial y^b} + {C^h}_{jb} {C^i}_{hc} -
-{C^h}_{jc} {C^i}_{hb} , $$
$${{S_b}^a}_{cd} =
 {\partial {C^a}_{bc} \over \partial y^d} -
{\partial {C^a}_{bd} \over \partial y^c} + {C^e}_{bc} {C^a}_{ed} - {C^e}_{bd}
{C^a}_{ec} .$$

The components of the Ricci d--tensor ${\bf R}_{\alpha \beta}$ with respect
 to locally adapted frame (3) are as follows:
\begin{equation}
R_{ij} = {{R_i}^k}_{jk}, R_{ia} =
 - {{}^2}P_{ia} = - {{P_i}^k}_{ka} , \label {f40}
\end{equation}
$$R_{ai} = {{}^1}P_{ai} = {{P_a}^b}_{ib} ,
 R_{ab} = S_{ab} = {{S_a}^c}_{bc} .$$
We point out that because, in general,  ${{}^1}P_{ai} \neq {{}^2}P_{ia}$ the
Ricci d--tensor is nonsymmetric.

The scalar curvature ${\bf R} = G^{\alpha \beta}{\bf R}_{\alpha \beta}$
is given by
\begin{equation}
{\bf R} = R + S ,   \label {f41}
\end{equation}
where $R = g^{ij} R_{ij} $ and $ S = h^{ab} S_{ab} .$

Now we can write the Einstein--Cartan equations for la--gravity
\begin{equation}
{\bf R}_{\alpha \beta } - {1\over 2} G_{\alpha \beta} {\bf R} + {\lambda}
G_{\alpha \beta} = {\kappa}_1 {\cal T}_{\alpha \beta} , \label {f42}
\end{equation}
and
\begin{equation}
{\bf T}^{\alpha}_{\cdot \beta \gamma} +
{G_{\beta}}^{\alpha} {\bf T}^{\tau}_{\cdot \gamma \tau} - {G_{\gamma}}^{\alpha}
{\bf T}^{\tau}_{\cdot \beta \tau}  =
 {\kappa}_2 {\cal Q}^{\alpha}_{\cdot \beta \gamma}, \label {f43}
\end{equation}
where ${\cal T}_{\alpha \beta}$ and ${\cal Q}^{\alpha}_{\cdot \beta \gamma}$
are respectively d--tensors of energy--momentum and spin--density of matter
on la--space $\xi ,$\quad ${\kappa}_1$ and ${\kappa}_2$ are corresponding
interaction constants and $\lambda$ is the cosmological constant. We have added the
algebraic system of equations for torsion (43) in order to close the system
of field equations (here we also point to the constraints (9)). We have
proposed a gauge like version of la--gravity for which a set of conservation
laws similar to those for Yang--Mills fields on la--spaces hold [12]. In
this work we restrict our considerations only to the
Einstein--Cartan la--gravity.

\subsection{Nonzero Divergence of the Energy--Momentum D--Tensor on LA--Spaces}
R.Miron and M.Anastasiei [1,2] pointed to this specific form of conservation
laws of matter on la--spaces:  They calculated the divergence of the
energy--momentum d--tensor on la--space $\xi,$
\begin{equation}
 D_{\alpha} {\cal T}^{\alpha}_{\beta} = {1\over \ {\kappa}_1} U_{\alpha} ,
\label {f44}
\end{equation}
and concluded that d--vector
$$ U_{\alpha} =
 {1\over 2} ( G^{\beta \delta}
 {{R_{\delta}}^{\gamma}}_{\phi \beta} {\bf T}^{\phi}_{\cdot \alpha \gamma} -
 G^{\beta \delta} {{R_{\delta}}^{\gamma}}_{\phi \alpha}
{\bf T}^{\phi}_{\cdot \beta \gamma} +
 {R^{\beta}_{\phi}} {\bf T}^{\phi}_{\cdot \beta \alpha} )$$
vanishes if and only if d--connection $D$ is without torsion.

 No wonder that conservation laws, in usual physical theories being a
 consequence of global (for usual gravity of local) automorphisms of the
fundamental space--time, are more sophisticated on the spaces with local
anisotropy. Here it is important to emphasize the multiconnection character
of la--spaces. For example, for a d--metric (8) on $\xi$ we can equivalently
introduce another (see (36)) metric linear connection $\tilde D$ with
coefficients
\begin{equation}
{\tilde {\Gamma}}^{\alpha}_{\cdot \beta \gamma} =
{1\over 2}  G^{\alpha \phi} ( {\delta G_{\beta \phi} \over \delta u^{\gamma}} +
{\delta G_{\gamma \phi} \over \delta u^{\alpha}} -
{\delta G_{\beta \gamma} \over \delta u^{\phi}}) . \label {f45}
\end{equation}
The Einstein equations
\begin{equation}
{\tilde {\bf R}}_{\alpha \beta} - {1\over 2} G_{\alpha \beta}
{\tilde {\bf R}} = {\kappa}_1 {\tilde {\cal T}}_{\alpha \beta}  \label {f46}
\end{equation}
constructed by using connection (43) have vanishing divergences
${\tilde D}^{\alpha} ( {{\tilde {\bf R}}_{\alpha \beta}} -
 {1\over 2} G_{\alpha \beta} {\tilde {\bf R}} ) = 0 $ and $ {\tilde D}^{\alpha}
{\tilde {\cal T}}_{\alpha \beta}=0 ,$ similarly as those on (pseudo)Riemannian
spaces. We conclude that by using the connection (43) we construct a model of
la--gravity which looks like locally isotropic on the total bundle $E.$
More general gravitational models with local anisotropy can be obtained by
using deformations of connection
 ${\tilde {\Gamma}}^{\alpha}_{\cdot \beta \gamma} ,$
\begin{equation}
 {{\Gamma}^{\alpha}}_{\beta \gamma} =
 {\tilde {\Gamma}}^{\alpha}_{\cdot \beta \gamma} + {P^{\alpha}}_{\beta \gamma}
+ {Q^{\alpha}}_{\beta \gamma} , \label {f47}
\end{equation}
were, for simplicity, ${{\Gamma}^{\alpha}}_{\beta \gamma}$ is chosen to be
also metric and satisfy Einstein equations (43).
We can consider deformation d--tensors ${P^{\alpha}}_{\beta \gamma}$ generated
(or not) by deformations of type (12) and (13) for na--maps. In this case
d--vector $U_{\alpha}$ can be interpreted as a generic source of local
anisotropy on $\xi$ satisfying generalized conservation laws (46).

\subsection{Deformation d--tensors and tensor--integral conservation laws }
>From (26) we obtain a tensor integral on    ${\cal C}({\xi})$  of a
d--tensor :
$$ N_{{\underline {\tau}}}^{{.}{\underline {\kappa}}}(\underline u)=
I_{{\overline S}_{p}}N^{{..}{\overline {\kappa}}}_
{{\overline {\tau}}{..}{\overline {\alpha}}_{1}{\ldots}{\overline {\alpha}}_{p}}
({\overline u})h^{{\overline {\tau}}}_{{\underline {\tau}}}({\underline u},
{\overline u})h^{{\underline {\kappa}}}_{{\overline {\kappa}}}
({\overline u}, {\underline u})dS^{{\overline {\alpha}}_{1}{\ldots}
{\overline {\alpha}}_{p}}.$$

We point out that tensor--integral can be defined not only for dc--tensors
but and for d--tensors on $\xi$. Really, suppressing indices ${\varphi}$~ and
${\gamma}$~ in (34) and (35), considering instead of a deformation dc--tensor
a deformation tensor
\begin{equation}
 B_{{\alpha}{\beta}}^{{..}{\gamma}}(u, {\underline u})=
B^{{..}{\gamma}}_{{\alpha}{\beta}}(u)=P^{\gamma}_{{.}{\alpha}{\beta}}(u)
 \label {f48}
\end{equation}
   (see deformations  (12) induced by a nc--transform) and integration
$ I_{S_{p}}{\ldots}dS^{{\alpha}_{1}{\ldots}{\alpha}_{p}}$
in la--space $\xi$ we obtain from (26) a tensor--integral on
$ {\cal C}({\xi})$~ of a d--tensor:
$$ N^{{.}{\underline {\kappa}}}_{{\underline {\tau}}}({\underline u})=
I_{S_{p}}N^{.{\kappa}}_{{\tau}{.}{\alpha}_{1}{\ldots}{\alpha}_{p}}(u)
h^{\tau}_{{\underline {\tau}}}({\underline u}, u)
h^{\underline {\kappa}}_{\kappa}(u, {\underline u})
dS^{{\alpha}_{1}{\ldots}{\alpha}_{p}}.$$
Taking into account (30) and using formulas (5) and (6) we can calculate
that curvature
$${\underline R}_{{\gamma}{.}{\alpha}{\beta}}^{.{\lambda}}=
D_{[{\beta}}B_{{\alpha}]{\gamma}}
^{{..}{\lambda}}+B^{{..}{\tau}}_{[{\alpha}{\vert}{\gamma}{\vert}}
B^{{..}{\lambda}}_{{\beta}]{\tau}}+
T^{{\tau}{..}}_{{.}{\alpha}{\beta}}
B^{{..}{\lambda}}_{{\tau}{\gamma}}$$
of connection ${\underline {\Gamma}}^{\gamma}_{{.}{\alpha}{\beta}}(u)=
{\Gamma}^{\gamma}_{{.}{\alpha}{\beta}}(u)+
B^{{..}{\gamma}}_{{\alpha}{\beta}{.}}(u), $
with $ B^{{..}{\gamma}}_{{\alpha}{\beta}}(u)$~
taken from (48), vanishes,
$ {\underline R}^{{.}{\lambda}}_{{\gamma}{.}{\alpha}{\beta}}=0. $
So, we can conclude that la--space $\xi$ admits a tensor integral structure
on ${\cal {C}}({\xi})$ for d--tensors
associated to deformation tensor $B^{{..}{\gamma}}_
{{\alpha}{\beta}}(u)$ if the nc--image  ${\underline {\xi}}$~ is
locally parallelizable. That way we generalize  the one space  tensor
integral constructions in [15,28,29], were the possibility  to introduce
tensor integral structure on a curved space was restricted by the condition that
 this space is locally parallelizable.
For $q=n+m$~  relations (35), written for d--tensor
${\cal N}^{{.}{\underline {\beta}}{\underline {\gamma}}}_
{\underline {\alpha}}$
(we change indices ${\overline {\alpha}}, {\overline {\beta}}, {\ldots}$ into
${\underline {\alpha}}, {\underline {\beta}}, {\ldots})$
extend the Gauss formula  on ${\cal {C}}({\xi})$:
\begin{equation}
 I_{S_{q-1}}{\cal N}_{\underline {\alpha}}^{{.}{\underline {\beta}}{\underline {\gamma}}}
d{\cal S}_{\underline {\gamma}}=I_{{\underline S}_{q}}{^{\underline {q-1}}
D}_{{\underline {\tau}}}{\cal N}_
{{\underline {\alpha}}}^{{.}{\underline {\beta}}{\underline {\tau}}}
d{\underline V}, \label {f49}
\end{equation}
where $d{\underline V}={\sqrt{{\vert}{\underline G}_{{\alpha}{\beta}}{\vert}}}
d{\underline u}^{1}{\ldots}d{\underline u}^{q}$ and
\begin{equation}
{^{\underline {q-1}}D}_{{\underline {\tau}}}{\cal N}_{\underline {\alpha}}^
{{.}{\underline {\beta}}{\underline {\tau}}}=
                      D_{{\underline {\tau}}}{\cal N}_{\underline {\alpha}}^
{{.}{\underline {\beta}}{\underline {\tau}}}-
T^{{\underline {\epsilon}}}_{{.}{\underline {\tau}}{\underline {\epsilon}}}
{\cal N}_{{\underline {\alpha}}}^{{\underline {\beta}}{\underline {\tau}}}-
B_{{\underline {\tau}}{\underline {\alpha}}}^{{..}{\underline {\epsilon}}}
{\cal N}_{{\underline {\epsilon}}}^{{.}{\underline {\beta}}{\underline {\tau}}}+
B_{{\underline {\tau}}{\underline {\epsilon}}}^{{..}{\underline {\beta}}}
{\cal N}_{{\underline {\alpha}}}^{{.}{\underline {\epsilon}}{\underline {\tau}}}.
\label {f50}
\end{equation}

Let consider physical values $N_{{\underline {\alpha}}}^{{.}{\underline {\beta}}}$
on ${\underline {\xi}}$~ defined on its density
${\cal N}_{{\underline {\alpha}}}^{{.}{\underline {\beta}}{\underline {\gamma}}}, $
i. e.
\begin{equation}
 N_{{\underline {\alpha}}}^{{.}{\underline {\beta}}}=
I_{{\underline S}_{q-1}}{\cal N}_{{\underline {\alpha}}}^{{.}{\underline {\beta}}
{\underline {\gamma}}}d{\cal S}_{{\underline {\gamma}}} \label {f51}
\end{equation}
with this conservation law  (due to (49)):
\begin{equation}
 {^{\underline {q-1}}D}_{{\underline {\gamma}}}{\cal N}_{{\underline {\alpha}}}
^{{.}{\underline {\beta}}{\underline {\gamma}}}=0. \label {f52}
\end{equation}
We note that these conservation laws differ from covariant conservation laws
for well known physical values such as density of electric current or of energy--
momentum tensor. For example, taking density ${\cal T}_{\beta}^{{.}{\gamma}},$
with corresponding to (50) and (52) conservation law,
\begin{equation}
{^{\underline {q-1}}D}_{{\underline {\gamma}}}{\cal T}_{{\underline {\beta}}}^
{{\underline {\gamma}}}=
D_{{\underline {\gamma}}}{\cal T}_{{\underline {\beta}}}^{{\underline {\gamma}}}
-T^{{\underline {\tau}}}_{{.}{\underline {\epsilon}} {\underline {\tau}}}
{\cal T}_{{\underline {\beta}}}^{{.}{\underline {\epsilon}}}-
B_{{\underline {\tau}}{\underline {\beta}}}^{{..}{\underline {\epsilon}}}
{\cal T}_{\underline {\epsilon}}^{{\underline {\tau}}}=0,  \label {f53}
\end{equation}
we can define values  (see (49) and (51))
$$ {\cal P}_{\alpha}=I_{{\underline S}_{q-1}}{\cal T}_{{\underline {\alpha}}}
^{{.}{\underline {\gamma}}}d{\cal S}_{{\underline {\gamma}}}. $$
Defined conservation laws (53) for ${\cal T}_{{\underline {\beta}}}^
{{.}{\underline {\epsilon}}}$ have nothing to do with those for energy--momentum
tensor $T_{\alpha}^{{.}{\gamma}} $  from Einstein equations for the almost
Hermitian gravity [1,2] or with ${\tilde {\bf T}}_{\alpha \beta}$
from (46) with vanishing divergence
$ D_{\gamma}{\tilde {\bf T}}_{\alpha}^{{.}{\gamma}}=0. $
So ${\tilde {\bf T}}_{\alpha}^{{.}{\gamma}}
{\neq}{\cal T}_{\alpha}^{{.}{\gamma}}.$ A similar
conclusion was made in [15] for unispacial locally isotropic tensor integral.
In the case of multispatial tensor integration we have another
possibility  (firstly pointed in [17,18] for Einstein-Cartan
spaces), namely, to identify ${\cal T}^{{.}{\underline {\gamma}}}
_{{\underline {\beta}}}$  from (53) with the na-image of
$ {\underline {\bf T}}^{{.}{\gamma}}_{\beta}$  on la--space $\xi.$ We shall
consider this construction in the next section.

\section {The Einstein Equations on Na--images of La--Spaces and
  Conservation Laws for La--Gravitational Fields}

It is well known that the standard pseudo--tensor description of the
energy--momentum values  for the Einstein gravitational fields is full of
ambiguities. Some light was shed by introducing additional geometrical structures
on curved space--time (bimetrics [35,36], biconnections [37], by taking into
account background spaces [38,34], or formulating variants of general relativity theory
on flat space [39,40]). We emphasize here that rigorous mathematical investigations
based on two (fundamental and background)
locally anisotropic, or isotropic, spaces should use well--defined,
motivated from physical point of view, mappings of these spaces. Our na--model
largely contains both attractive features of the mentioned approaches; na--maps
establish a local 1--1 correspondence between the fundamental la--space
and auxiliary la--spaces on which biconnection (or even multiconnection)
structures are induced. But these structures are not a priory postulated as in
a lot of gravitational theories, we tend to specify them to be locally
reductible  to the locally isotropic Einstein theory [38,30].

Let us consider a fixed background la--space $\underline {\xi}$ with given metric
${\underline G}_{\alpha \beta} = ({\underline g}_{ij} , {\underline h}_{ab} )$
and d--connection ${\underline {\tilde {\Gamma}}}^{\alpha}_{\cdot \beta\gamma},$
for simplicity in this subsection we consider compatible metric and connections
 being torsionless and with vanishing curvatures.
Supposing that there is an nc--transform from the fundamental  la--space
$\xi$ to the auxiliary $\underline {\xi} .$ we are interested  in the equivalents
of the Einstein equations (46) on $\underline {\xi} .$

We consider that a part of gravitational degrees of freedom is "pumped out" into
the dynamics of deformation d--tensors for d--connection,
  ${P^{\alpha}}_{\beta \gamma},$ and metric, $B^{\alpha \beta} =
 ( b^{ij} , b^{ab} ) .$ The remained part of degrees of freedom  is coded into
the metric ${\underline G}_{\alpha \beta}$ and d--connection
${\underline {\tilde {\Gamma}}}^{\alpha}_{\cdot \beta \gamma} .$

Following [38,30] we apply the first order formalism and consider $B^{\alpha \beta}$
and ${P^{\alpha}}_{\beta \gamma}$ as independent variables on $\underline {\xi}.$
Using notations $$P_{\alpha} = {P^{\beta}}_{\beta \alpha} ,\quad
{\Gamma}_{\alpha} = {{\Gamma}^{\beta}}_{\beta \alpha} ,$$
$$
 {\hat B}^{\alpha \beta} = \sqrt{|G|}  B^{\alpha \beta} ,
 {\hat G}^{\alpha \beta} = \sqrt{|G|}  G^{\alpha \beta} ,
{\underline {\hat G}}^{\alpha \beta} = \sqrt{|\underline G |}
{\underline G}^{\alpha \beta}$$ and making identifications
$${\hat B}^{\alpha \beta} + {\underline {\hat G}}^{\alpha \beta} =
 {\hat G}^{\alpha \beta}, {\quad}{\underline {\Gamma}}^{\alpha}_{\cdot \beta \gamma}
 - {P^{\alpha}}_{\beta \gamma} = {{\Gamma}^{\alpha}}_{\beta \gamma},$$
we take the action of la--gravitational field on $\underline {\xi}$ in this
form:
\begin{equation}
{\underline {\cal S}}^{(g)} = - {(2c{\kappa}_1  )}^{-1} \int {\delta}^{q} u {}
{\underline {\cal L}}^{(g)} , \label {f54}
\end{equation}
where
$${\underline {\cal L}}^{(g)} = {\hat B}^{\alpha \beta} ( D_{\beta} P_{\alpha} -
 D_{\tau} {P^{\tau}}_{\alpha \beta} ) +$$
$$ ( {\underline {\hat G}}^{\alpha \beta} +
 {\hat B}^{\alpha \beta} ) ( P_{\tau} {P^{\tau}}_{\alpha \beta} -
 {P^{\alpha}}_{\alpha \kappa} {P^{\kappa}}_{\beta \tau} )$$
and the interaction constant is taken ${\kappa}_1 = {4{\pi}\over {c^4}} k,{\quad}
(c$ is the light constant and $k$ is Newton constant) in order to obtain
concordance with the Einstein theory in the locally isotropic limit.

We construct on $\underline {\xi}$ a la--gravitational theory with matter fields
(denoted as ${\varphi}_A$ with $A$ being a general index) interactions by postulating
this Lagrangian density for matter fields
\begin{equation}
{\underline {\cal L}}^{(m)} = {\underline {\cal L}}^{(m)}
 [{\underline {\hat G}}^{\alpha \beta} + {\hat B}^{\alpha \beta} ;
{\delta \over \delta  u^{\gamma}}
 ( {\underline {\hat G}}^{\alpha \beta} + {\hat B}^{\alpha \beta} ) ;
 {\varphi}_A ; {\delta {\varphi}_A \over \delta u^{\tau} }]. \label {f55}
\end{equation}

Starting from (54) and (55) the total action of la--gravity on $\underline {\xi}$
 is written as
\begin{equation}
{\underline {\cal S}} = {(2c {\kappa}_1 )}^{-1}
\int {\delta}^q u {\underline {\cal L}}^{(g)} +
 c^{-1} \int {\delta}^{(m)} {\underline {\cal L}}^{(m)} . \label {f56}
\end{equation}
Applying variational procedure on $\underline {\xi} ,$ similar to that presented
in [38] but in our case adapted to N--connection by using derivations (3)
instead of partial derivations, we derive from (56) the la--gravitational
field equations
\begin{equation}
{\bf {\Theta}}_{\alpha \beta} =
 {{\kappa}_1} ( {\underline {\bf t}}_{\alpha \beta} +
{\underline {\bf T}}_{\alpha \beta} ) \label {f57}
\end{equation}
and matter field equations
\begin{equation}
{{\triangle}{\underline {\cal L}}^{(m)} \over \triangle
 {\varphi}_A } = 0 , \label {f58}
\end{equation}
where $\triangle \over \triangle {\varphi}_A$ denotes the variational derivation.

In (57) we have introduced  these values:
the energy--momentum d--tensor for la--gravitational field
\widetext
\begin{equation}
{\kappa}_1 {\underline {\bf t}}_{\alpha \beta} =
 ({\sqrt{|G|}})^{-1} {\triangle {\underline {\cal L}}^{(g)} \over
\triangle G^{\alpha \beta}} =
K_{\alpha \beta} + {P^{\gamma}}_{\alpha \beta} P_{\gamma} -
{P^{\gamma}}_{\alpha \tau} {P^{\tau}}_{\beta \gamma} +
{1\over 2} {\underline G}_{\alpha \beta} {\underline G}^{\gamma \tau}
 ( {P^{\phi}}_{\gamma \tau} P_{\phi} - {P^{\phi}}_{\gamma \epsilon}
 {P^{\epsilon}}_{\phi \tau} ),  \label {f59}
\end{equation}
(where $$K_{\alpha \beta} = {\underline D}_{\gamma}  K^{\gamma}_{\alpha \beta} ,$$
$$ 2K^{\gamma}_{\alpha \beta} = - B^{\tau \gamma} {P^{\epsilon}}_{\tau ( \alpha}
{\underline G}_{\beta ) \epsilon} - B^{\tau \epsilon}
 {P^{\gamma}}_{\epsilon ( \alpha} {\underline G}_{\beta ) \tau} +$$
$$ {\underline G}^{\gamma \epsilon} h_{\epsilon ( \alpha} P_{\beta )} +
{\underline G}^{\gamma \tau} {\underline G}^{\epsilon \phi}
 {P^{\varphi}}_{\phi \tau} {\underline G}_{\varphi ( \alpha} B_{\beta ) \epsilon}
 + {\underline G}_{\alpha \beta} B^{\tau \epsilon} {P^{\gamma}}_{\tau \epsilon} -
 B_{\alpha \beta} P^{\gamma}{\quad}) ,$$
$$2{\bf \Theta} = {\underline D}^{\tau} {\underline D}_{tau} B_{\alpha \beta} +
{\underline G}_{\alpha \beta} {\underline D}^{\tau} {\underline D}^{\epsilon}
B_{\tau \epsilon} - {\underline G}^{\tau \epsilon} {\underline D}_{\epsilon}
{\underline D}_{(\alpha} B_{\beta ) \tau} $$
and the energy--momentum d--tensor of matter
\begin{equation}
{\underline {\bf T}}_{\alpha \beta} = 2 {\triangle {\cal L}^{(m)} \over
 \triangle {\underline {\hat G}}^{\alpha \beta}} -
  {\underline G}_{\alpha \beta}
 {\underline G}^{\gamma \delta} {\triangle {\cal L}^{(m)} \over \triangle
{\underline {\hat G}}^{\gamma \delta}} .\label {f60}
\end{equation}
As a consequence of (58)--(60) we obtain the d--covariant on $\underline {\xi}$
conservation laws
\begin{equation}
 {\underline D}_{\alpha} ( {\underline {\bf t}}^{\alpha \beta} +
{\underline {\bf T}}^{\alpha \beta} ) = 0. \label {f61}
\end{equation}
We have postulated the Lagrangian density of matter fields (55) in a form as to
treat ${\underline {\bf t}}^{\alpha \beta} + {\underline {\bf T}}^{\alpha \beta}$
as the source in (57).

Now we formulate the main results of this section:
\begin{prop}
 The dynamics of the Einstein la--gravitational fields,
modeled as solutions of equations (46) and
matter fields on la--space $\xi ,$ can be equivalently locally modeled on
a background  la--space $\underline {\xi}$ provided with a trivial d-connection
and metric structures having zero d--tensors of torsion and curvature by field
equations (57) and (58) on condition that deformation tensor
 ${P^{\alpha}}_{\beta \gamma}$ is a solution of the Cauchy problem posed for
basic equations for a chain of na--maps from $\xi$ to $\underline {\xi}.$
\end{prop}

\begin{prop}
 Local, d--tensor, conservation laws for Einstein
la--gravitational fields can be written in form (61)  for
la--gravitational (59) and matter (60) energy--momentum d--tensors. These laws
are d--covariant on the background space $\underline {\xi}$ and must be completed
with invariant conditions of type (22)-(25) for every deformation parameters
of a chain of na--maps from $\xi$ to $\underline {\xi}.$
\end{prop}

The above presented considerations represent the proofs of both propositions.

We emphasize that nonlocalization  of both locally
anisotropic and isotropic gravitational energy--momentum values on the
 fundamental \quad (locally anisotropic or isotropic)\quad  space \quad $\xi$
 \quad is a consequence of the absence of global group
automorphisms for generic curved spaces. Considering gravitational theories
from view of multispaces and their mutual maps {\quad} (directed by the basic
geometric structures  on $\xi \quad$ such as {\quad} N--connection, d--connection, d--torsion
and d--curvature components, see coefficients for basic na--equations
(19)-(21)),{\quad} we can formulate local d--tensor  conservation laws on auxiliary
globally automorphic spaces being related with space $\xi$ by means of chains
of na--maps. Finally, we remark that as a matter of principle we can use
d--connection deformations of type (47) in order to modelate the la--gravitational
interactions with nonvanishing torsion and nonmetricity. In this case we must
introduce a corresponding source in (61) and define generalized conservation
laws as in (44) \quad (see similar details for locally isotropic generalizations of
the Einstein gravity in Refs [16,17,21]).

\section{OUTLOOK AND CONCLUSIONS}

In this paper we have presented a detailed study of the problem of formulation
of conservation laws on spaces with local anisotropy. The need for such an
investigation was often expressed in order to develop a number of geometrical
models of interactions in locally anisotropic media or to extend in a
self--consistent manner some gravitational and matter field  theories on
tangent and vector bundles. As a geometric background of our considerations
we have chosen the R.Miron and M.Anastasiei [1,2] general approach of modelling
la--spaces on vector bundles provided with compatible nonlinear and distinguished
connections and metric structures. This allowed us to obtain a similarity,
within certain limits,  with the Einstein--Cartan spaces  and to generalize
some our results  on both locally anisotropic and isotropic gravitational
theories [11,12,16-21,24-30].

We have formulated the theory of nearly autoparallel maps of la--spaces. Such
maps generalize the conformal transforms of curved spaces, are characterized
by similar invariants as Weyl tensor and Thomas parameters and appear to have
a lot of applications in modern classical and quantum gravity. We have shown
that, as a matter of principle, we can modelate locally equivalent physical
theories on every la--spaces interrelated with the fundamental space--time
by mean of na--maps. We can introduce a new classification of la-spaces with
respect to nearly conformal transforms to background one with trivial connections
and vanishing torsions  and curvatures. On such backgrounds admitting global
on corresponding vector bundles, but locally anisotropic on base space, groups
 of automorphisms we can define in a usual manner, for a chosen model of field
 la--interactions, the conservation laws. We must complete these laws by a set
of invariant conditions being associated to chains of na--maps.

The main advantage of our geometric constructions is theirs compatibility
with similar ones introduced in the framework of the tensor integral formalism.
We have introduced bi-- and multitensors and defined nearly autoparallel
tensor integral on bi-- and multispaces. We have formulated corresponding
conservation laws on la--multispaces.

Our results points the possibility of formulation of physical models of locally
anisotropic field interactions  and definition of conservation laws on la--spaces
in spite of the fact that,
 at a glance, the generic local anisotropy of such spaces cast doubt on the
general possibility of formulation of such problems.

\acknowledgments
The author is very grateful to Academician R. Miron and Professor M. Anastasiei for
discussion and comments.
\appendix
\section
{Proof of the Theorem 2}
We sketch the proof respectively  for every point in the theorem:
\begin{enumerate}
\item  It is easy to verify that a--parallel equations (14) on $\xi$
transform into similar ones on $\underline{\xi}$ if and only if deformations
(12) with deformation d--tensors of type ${P^{\alpha}}_{\beta \gamma} (u) =
 {\psi}_{( \beta} {\delta}^{\alpha}_{\gamma )}$ are considered.
\item Using corresponding to $na_{(1)}$--maps parametrizations of $a(u,v)$ and
$b(u,v)$ (see conditions of the theorem) for arbitrary $v^{\alpha}\neq 0$ on
$U \in \xi$ and after a redefinition of deformation parameters we obtain that
equations (16) hold if and only if ${P^{\alpha}}_{\beta \gamma}$ satisfies (13).
\item  In a similar manner we obtain basic $na_{(2)}$--map equations  (20) from
(16) by considering $na_{(2)}$--parametrizations of deformation parameters and
d--tensor.
\item  For $na_{(3)}$--maps we mast take into consideration deformations of
torsion (13) and introduce  $na_{(3)}$--parametrizations for $b(u,v)$ and
${P^{\alpha}}_{\beta \gamma}$ into the basic na--equations (16). The last ones
for $na_{(3)}$--maps are equivalent to equations (21) (with a corresponding
 redefinition of deformation parameters).
\end{enumerate}
\section
{Proof of the Theorem 3}
\begin{enumerate}
\item  Let us prove that a--invariant conditions (22) hold. Deformations of d--connections
of type \begin{equation}
{}^{(0)}{\underline \gamma}^{\mu}_{\cdot \alpha \beta} =
 {{\gamma}^{\mu}}_{\alpha \beta} +
 {\psi}_{( \alpha} {\delta}^{\mu}_{\beta )} \label {f b1}
\end{equation}
define a--applications. Contracting indices  $\mu$ and $\beta$ we can write
\begin{equation}
{\psi}_{\alpha} =
 {1\over m+n+1} ( {{\underline {\gamma}}^{\beta}}_{\alpha \beta} -
 {{\gamma}^{\beta}}_{\alpha \beta} ) . \label {f b2}
\end{equation}
Introducing d--vector ${\psi}_{\alpha}$ into previous relation and expressing
$${{\gamma}^{\alpha}}_{\beta \tau} = - {T^{\alpha}}_{\beta \tau}  +
 {{\Gamma}^{\alpha}}_{\beta \tau} $$
and similarly for underlined values we obtain the first invariant conditions
from (22).

Putting deformation (B1) into the formula for
$${\underline r}^{\cdot \tau}_{\alpha \cdot \beta \gamma} \quad \mbox{and} \quad
{\underline r}_{\alpha \beta} =
 {\underline r}^{\cdot \tau}_{\alpha \tau \beta \tau}$$
we obtain respectively  relations
\widetext

\begin{equation}
{\underline r}^{\cdot \tau}_{\alpha \cdot \beta \gamma} -
 r^{\cdot \tau}_{\alpha \cdot \beta \gamma} =
 {\delta}^{\tau}_{\alpha} {\psi}_{[ \gamma \beta ]} +
 {\psi}_{\alpha [ \beta} {\delta}^{\tau}_{\gamma ]} +
 {\delta}^{\tau}_{( \alpha} {\psi}_{\varphi )} {w^{\varphi}}_{\beta \gamma}
\label {f b3}  \end{equation}
and
\begin{equation}
{\underline r}_{\alpha \beta} - r_{\alpha \beta} =
 {\psi}_{[ \alpha \beta ]} +
 (n+m-1) {\psi}_{\alpha \beta} + {\psi}_{\varphi} {w^{\varphi}}_{\beta \alpha} +
 {\psi}_{\alpha} {w^{\varphi}}_{\beta \varphi},  \label {f b4} \end{equation}
where
$${\psi}_{\alpha \beta} = {}^{({\gamma})}D_{\beta} {\psi}_{\alpha} -
{\psi}_{\alpha} {\psi}_{\beta} .$$
Putting (B1) into  (B3) we  can express ${\psi}_{[ \alpha \beta ]}$ as

$$
{\psi}_{[ \alpha \beta ]} = {1\over n+m+1} [ {\underline r}_{[ \alpha \beta]} +
{2\over n+m+1} {\underline {\gamma}}^{\tau}_{\cdot \varphi \tau}
 {w^{\varphi}}_{[ \alpha \beta ]} -
{1\over n+m+1} {\underline {\gamma}}^{\tau}_{\cdot \tau [ \alpha}
 {w^{\varphi}}_{\beta ] \varphi} ]- $$
\begin{equation}
{1\over n+m+1} [ r_{[ \alpha \beta ]} +
{2\over n+m+1} {{\gamma}^{\tau}}_{\varphi \tau} {w^{\varphi}}_{[ \alpha \beta ]} -
 {1\over n+m+1} {{\gamma}^{\tau}}_{\tau [ \alpha} {w^{\varphi}}_{\beta ] \varphi} ] .
 \label {f b5}  \end{equation}
To simplify our consideration we can choose an  a--transform, parametrized by
corresponding $\psi$--vector from (B1), (or fix a local coordinate cart) the
antisymmetrized relations (B5) to be satisfied by  d--tensor
$$
 {\psi}_{\alpha \beta} = {1\over n+m+1} [ {\underline r}_{\alpha \beta} +
 {2\over n+m+1} {\underline {\gamma}}^{\tau}_{\cdot \varphi \tau}
{w^{\varphi}}_{\alpha \beta} -
 {1\over n+m+1} {\underline {\gamma}}^{\tau}_{\cdot \alpha \tau}
 {w^{\varphi}}_{\beta \varphi} - r_{\alpha \beta} -$$
\begin{equation}
{2\over n+m+1} {{\gamma}^{\tau}}_{\varphi \tau} {w^{\varphi}}_{\alpha \beta} +
{1\over n+m+1} {{\gamma}^{\tau}}_{\alpha \tau} {w^{\varphi}}_{\beta \varphi} ]
\label {f b6}  \end{equation}
Introducing expressions (B1),(B5) and (B6) into  deformation of curvature (B2)
we obtain the second conditions (22) of a-map invariance:
$$^{(0)}W^{\cdot \delta}_{\alpha \cdot \beta \gamma} =
{}^{(0)} {\underline W}^{\cdot \delta}_{\alpha \cdot \beta \gamma} , $$
where the Weyl d--tensor on $\underline {\xi}$ (the extension of the usual one
 for geodesic maps on (pseudo)--Riemannian spaces to the case of v--bundles
 provided with N--connection structure) is defined as
$${}^{(0)}{\underline W}^{\cdot \tau}_{\alpha \cdot \beta \gamma} =
   {\underline r}^{\cdot \tau}_{\alpha \cdot \beta \gamma} +
{1\over n+m+1} [ {\underline {\gamma}}^{\tau}_{\cdot \varphi \tau}
 {\delta}^{\tau}_{( \alpha} {w^{\varphi}}_{\beta ) \gamma} -
( {\delta}^{\tau}_{\alpha}{\underline r}_{[ \gamma \beta ]}  +
{\delta}^{\tau}_{\gamma} {\underline r}_{[ \alpha \beta ]} -
{\delta}^{\tau}_{\beta} {\underline r}_{[ \alpha \gamma ]} )] -$$
$$
{1\over {(n+m+1)}^2} [ {\delta}^{\tau}_{\alpha}
 ( 2 {\underline {\gamma}}^{\tau}_{\cdot \varphi \tau}
 {w^{\varphi}}_{[ \gamma \beta ] } -
{\underline {\gamma}}^{\tau}_{\cdot \tau [ \gamma }
{w^{\varphi}}_{\beta ] \varphi} ) + {\delta}^{\tau}_{\gamma}
 ( 2 {\underline {\gamma}}^{\tau}_{\cdot \varphi \tau}
  {w^{\varphi}}_{\alpha \beta} -{\underline {\gamma}}^{\tau}_{\cdot \alpha \tau}
{w^{\varphi}}_{\beta \varphi}) -$$   $$ {\delta}^{\tau}_{\beta}
 ( 2 {\underline {\gamma}}^{\tau}_{\cdot \varphi \tau}
  {w^{\varphi}}_{\alpha \gamma} - {\underline {\gamma}}^{\tau}_{\cdot \alpha \tau}
{w^{\varphi}}_{\gamma \varphi} ) ]$$
The formula for $^{(0)}W^{\cdot \tau}_{\alpha \cdot \beta \gamma}$
written similarly with respect to non--underlined values is presented in
Section IV.
\item
To obtain $na_{(1)}$--invariant conditions we rewrite $na_{(1)}$--equations
(19) as to consider in explicit form covariant derivation $^{({\gamma})}D$ and
deformations (12) and (13):
$$
2( {}^{({\gamma})}D_{\alpha} {P^{\delta}}_{\beta \gamma} +
 {}^{({\gamma})}D_{\beta} {P^{\delta}}_{\alpha \gamma} +
 {}^{({\gamma})}D_{\gamma} {P^{\delta}}_{\alpha \beta} +
{P^{\delta}}_{\tau \alpha} {P^{\tau}}_{\beta \gamma} +
{P^{\delta}}_{\tau \beta}  {P^{\tau}}_{\alpha \gamma} +
{P^{\delta}}_{\tau \gamma} {P^{\tau}}_{\alpha \beta} ) =$$
\begin{equation}
{T^{\delta}}_{\tau ( \alpha} {P^{\tau}}_{\beta \gamma )} +
{H^{\delta}}_{\tau ( \alpha} {P^{\tau}}_{\beta \gamma )} +
b_{( \alpha} {P^{\delta}}_{\beta \gamma )} + a_{( \alpha \beta}
 {\delta}^{\delta}_{\gamma )}. \label {f b7}  \end{equation}
Alternating the first two indices in (B7) we have
$$2 ({\underline r}^{\cdot \delta}_{( \alpha \cdot \beta ) \gamma} -
              r^{\cdot \delta}_{( \alpha \cdot \beta ) \gamma}) =$$
$$
 2( {}^{(\gamma )}D_{\alpha} {P^{\delta}}_{\beta \gamma} +
   {}^{(\gamma )}D_{\beta} {P^{\delta}}_{\alpha \gamma} -
  2{}^{(\gamma )}D_{\gamma} {P^{\delta}}_{\alpha \beta} +
 {P^{\delta}}_{\tau \alpha} {P^{\tau}}_{\beta \gamma} +
   {P^{\delta}}_{\tau \beta} {P^{\tau}}_{\alpha \gamma} -
  2{P^{\delta}}_{\tau \gamma} {P^{\tau}}_{\alpha \beta} ) .$$
Substituting the last expression from (B7) and rescalling the
deformation parameters and d--tensors we obtain the conditions
(19).
\item
Now we prove the invariant conditions for $na_{(0)}$--maps satisfying
conditions $$ \epsilon \neq 0 \quad \mbox{and} \quad \epsilon -
F^{\alpha}_{\beta} F^{\beta}_{\alpha} \neq 0 $$
Let define  the auxiliary d--connection
\begin{equation}
{\tilde {\gamma}}^{\alpha}_{\cdot \beta \tau} =
{\underline {\gamma}}^{\alpha}_{\cdot \beta \tau}  - {\psi}_{( \beta}
{\delta}^{\alpha}_{\tau )} = {{\gamma}^{\alpha}}_{\beta \tau} +
{\sigma}_{( \beta} F^{\alpha}_{\tau )}  \label {f b8} \end{equation}
and write
$$ {\tilde D}_{\gamma} = {}^{({\gamma})}D_{\gamma} F^{\alpha}_{\beta} +
{\tilde {\sigma}}_{\gamma} F^{\alpha}_{\beta} - {\epsilon} {\sigma}_{\beta}
{\delta}^{\alpha}_{\gamma} ,$$
where $ {\tilde {\sigma}}_{\beta} = {\sigma}_{\alpha} F^{\alpha}_{\beta}, $
or, as a consequence from the last equality,
$${\sigma}_{(\alpha} F^{\tau}_{\beta )} =
{\epsilon} F^{\tau}_{\lambda} ( {}^{({\gamma})}D_{(\alpha} F^{\alpha}_{\beta )}
- {\tilde D}_{( \alpha} F^{\lambda}_{\beta )} ) +
 {\tilde {\sigma}}_({\alpha} {\delta}^{\tau}_{\beta )}.$$
Introducing auxiliary connections
$$ {\star {\gamma}}^{\alpha}_{\cdot \beta \lambda} =
          {\gamma}^{\alpha}_{\cdot \beta \lambda} + {\epsilon} F^{\alpha}_{\tau}
{}^{({\gamma})}D_{(\beta} F^{\tau}_{\lambda )} $$ and
$$
{\check {\gamma}}^{\alpha}_{\cdot \beta \lambda} =
{\tilde {\gamma}}^{\alpha}_{\cdot \beta \lambda} + {\epsilon} F^{\alpha}_{\tau}
{\tilde D}_{(\beta} F^{\tau}_{\lambda )}$$
we can express deformation (B8) in a form
characteristic for a--maps:
\begin{equation}
{\hat {\gamma}}^{\alpha}_{\cdot \beta \gamma} =
 {\star {\gamma}}^{\alpha}_{\cdot \beta  \gamma} +
{\tilde {\sigma}}_{( \beta} {\delta}^{\alpha}_{\lambda )} . \label {f b9}
\end{equation}
Now it's obvious that $na_{(2)}$--invariant conditions (24) are equivalent
with a--invariant conditions (22) written for d--connection (B9). As a matter
of principle we can write formulas for such $na_{(2)}$--invariants in terms
of "underlined" and "non--underlined" values by expressing consequently all used
 auxiliary connections as deformations of "prime" connections on $\xi$ and
"final" connections on $\underline {\xi}.$ We omit such tedious calculations
in this work.
\item
Finally, we prove the last statement, for $na_{(3)}$--maps,  of the theorem 3.
Let
\begin{equation}
q_{\alpha} {\varphi}^{\alpha} = e = \pm 1 , \label {f b10}
\end{equation}
where ${\varphi}^{\alpha}$ is contained in
\begin{equation}
{\underline {\gamma}}^{\alpha}_{\cdot \beta \gamma} =
{{\gamma}^{\alpha}}_{\beta \gamma} + {\psi}_{(\beta} {\delta}^{\alpha}_{\gamma )} +
{\sigma}_{\beta \gamma} {\varphi}^{\alpha} . \label {f b11}
\end{equation}
Acting with operator $^{({\gamma})}{\underline D}_{\beta}$ on (B10) we write
\begin{equation}
{}^{({\gamma})} {\underline D}_{\beta} q_{\alpha} = {}^{({\gamma})} D_{\beta}
 q_{\alpha} - {\psi}_{(\alpha} q_{\beta )} - e {\sigma}_{\alpha \beta}.
\label {f b12} \end{equation}
Contracting (B12) with ${\varphi}^{\alpha}$ we can express
$$ e{\varphi}^{\alpha} {\sigma}_{\alpha \beta} = {\varphi}^{\alpha}
 ({}^{({\gamma})} D_{\beta} q_{\alpha} -
 {}^{({\gamma})}{\underline D}_{\beta} q_{\alpha} ) -
{\varphi}_{\alpha} q^{\alpha} q_{\beta} - e {\psi}_{\beta} .$$
Putting the last formula in (B11) contracted on indices $\alpha$ and $\gamma$
we obtain
\widetext

\begin{equation}
(n+m){\psi}_{\beta} = {\underline {\gamma}}^{\alpha}_{\cdot \alpha \beta} -
{{\gamma}^{\alpha}}_{\alpha \beta} +
 e{\psi}_{\alpha} {\varphi}^{\alpha} q_{\beta} +
e {\varphi}^{\alpha} {\varphi}^{\beta} ( {}^{({\gamma})}{\underline D}_{\beta} -
{}^{({\gamma})}D_{\beta}  ) . \label {f b13} \end{equation}
>From these relations, taking into consideration (B10), we have

\widetext

$$ (n+m-1) {\psi}_{\alpha} {\varphi}^{\alpha} =$$
$$ {\varphi}^{\alpha} ( {\underline {\gamma}}^{\alpha}_{\cdot \alpha \beta} -
{{\gamma}^{\alpha}}_{\alpha \beta} ) + e {\varphi}^{\alpha} {\varphi}^{\beta}
({}^{({\gamma})}{\underline D}_{\beta}q_{\alpha} -
 {}^{({\gamma})} D_{\beta}q_{\alpha} ) $$

Using the equalities and identities (B12) and (B13) we can express deformations
(B11) as the first $na_{(3)}$--invariant conditions from (25).
\par
To prove the second class of $na_{(3)}$--invariant conditions we introduce
two additional d--tensors:
\begin{equation}
{{\rho}^{\alpha}}_{\beta \gamma \delta} =
r^{\cdot \alpha}_{\beta \cdot \gamma \delta}  +
{1\over 2}( {\psi}_{(\beta} {\delta}^{\alpha}_{\varphi )} +
 {\sigma}_{\beta \varphi}{\varphi}^{\tau} ) {w^{\varphi}}_{\gamma \delta}{\quad}
 \mbox{and} {\quad}
{\underline {\rho}}^{\alpha}_{\cdot \beta \gamma \delta} =
{\underline r}^{\cdot \alpha}_{\beta \cdot \gamma \delta} -
{1\over 2}( {\psi}_{(\beta} {\delta}^{\alpha}_{{\varphi})} -
 {\sigma}_{\beta \varphi}{\varphi}^{\tau} ) {w^{\varphi}}_{\gamma \delta} .
\label {f b14} \end{equation}
Using deformation (B11) and (B14) we write relation
\begin{equation}
{\tilde {\sigma}}^{\alpha}_{\cdot \beta \gamma \delta} =
{\underline {\rho}}^{\alpha}_{\cdot \beta \gamma \delta} -
{\rho}^{\alpha}_{\cdot \beta \gamma \delta} =
{\psi}_{\beta [\delta} {\delta}^{\alpha}_{\gamma ]} -
 {\psi}_{[{\gamma}{\delta}]} {\delta}^{\alpha}_{\beta} -
 {\sigma}_{\beta \gamma \delta} {\varphi}^{\alpha} ,
\label {f b15}  \end{equation}
where
$${\psi}_{\alpha \beta} = {}^{({\gamma})}D_{\beta} {\psi}_{\alpha} +
{\psi}_{\alpha} {\psi}_{\beta} - ({\nu} + {\varphi}^{\tau} {\psi}_{\tau} )
 {\sigma}_{\alpha \beta} ,$$
and
$$
{\sigma}_{\alpha \beta \gamma} = {}^{({\gamma})}D_{[\gamma}
{\sigma}_{{\beta}]{\alpha}} + {\mu}_{[\gamma} {\sigma}_{{\beta}]{\alpha}}
 - {\sigma}_{{\alpha}[{\gamma}} {\sigma}_{{\beta}]{\tau}} {\varphi}^{\tau} .$$
Let multiply (B15) on $q_{\alpha}$ and write (taking into account relations
(B10)) the relation
\begin{equation}
e{\sigma}_{\alpha \beta \gamma} = - q_{\tau}
 {\tilde{\sigma}}^{\tau}_{\cdot \alpha \beta \delta} +
 {\psi}_{\alpha [ \beta} q_{\gamma ]} - {\psi}_{[\beta \gamma ]} q_{\alpha} .
\label {f B16} \end{equation}
The next step is to express ${\psi}_{\alpha \beta}$ trough d--objects on ${\xi}.$
To do this we contract indices $\alpha$ and $\beta$ in (B15) and obtain
$$(n+m) {\psi}_{[\alpha \beta ]} = - {\sigma}^{\tau}_{\cdot \tau \alpha \beta} +
 eq_{\tau}{\varphi}^{\lambda} {\sigma}^{\tau}_{\cdot \lambda \alpha \beta} -
 e{\tilde {\psi}}_{[\alpha} {\tilde {\psi}}_{\beta ]} .$$
Then contracting indices $\alpha$ and $\delta$ in (B15) and using (B16) we
write
\begin{equation}
(n+m-2){\psi}_{\alpha \beta} =
{\tilde {\sigma}}^{\tau}_{\cdot \alpha \beta \tau} -
eq_{\tau}{\varphi}^{\lambda} {\tilde {\sigma}}^{\tau}_{\cdot \alpha \beta \lambda}
+ {\psi}_{[ \beta \alpha ]} + e( {\tilde {\psi}}_{\beta} q_{\alpha} -
{\hat {\psi}}_{(\alpha} q_{\beta )} ,
\label {f b17} \end{equation}
where ${\hat{\psi}}_{\alpha} = {\varphi}^{\tau} {\psi}_{\alpha \tau}.$
If the both parts of (B17) are contracted with ${\varphi}^{\alpha} ,$
it results that
$$(n+m-2){\tilde {\psi}}_{\alpha} =
 {\varphi}^{\tau}{\sigma}^{\lambda}_{\cdot \tau \alpha \lambda} -
e q_{\tau} {\varphi}^{\lambda} {\varphi}^{\delta}
{\sigma}^{\tau}_{\lambda \alpha \delta} - eq_{\alpha} ,$$
and, in consequence of ${\sigma}^{\alpha}_{\beta ( \gamma \delta )} = 0,$
we have $$ (n+m-1){\varphi} = {\varphi}^{\beta}
 {\varphi}^{\gamma} {\sigma}^{\alpha}_{\cdot \beta \gamma \alpha} .$$
By using the last expressions we can write
\begin{equation}
(n+m-2) {\underline {\psi}}_{\alpha} =
{\varphi}^{\tau} {\sigma}^{\lambda}_{\cdot \tau \alpha \lambda} -
eq_{\tau} {\varphi}^{\lambda}{\varphi}^{\delta}
 {\sigma}^{\tau}_{\cdot \lambda \alpha \delta} - e {(n+m-1)}^{-1} q_{\alpha}
{\varphi}^{\tau}{\varphi}^{\lambda}
 {\sigma}^{\delta}_{\cdot \tau \lambda \delta} . \label{f b18} \end{equation}
Contracting (B17) with ${\varphi}^{\beta}$ we have
$$(n+m){\hat{\psi}}_{\alpha} = {\varphi}^{\tau}
{\sigma}^{\lambda}_{\cdot  \alpha \tau \lambda} + {\tilde{\psi}}_{\alpha}$$
and taking into consideration (B18) we can express ${\hat {\psi}}_{\alpha}$
through ${\sigma}^{\alpha}_{\cdot \beta \gamma \delta}.$

 As a consequence of (B16)--(B18) we obtain this formulas for d--tensor
${\psi}_{\alpha \beta} :$
\widetext
$$
(n+m-2){\psi}_{\alpha \beta} = {\sigma}^{\tau}_{\cdot \alpha \beta \tau} -
eq_{\tau} {\varphi}^{\lambda} {\sigma}^{\tau}_{\cdot \alpha \beta \lambda} +$$
$${1\over n+m} \{ -{\sigma}^{\tau}_{\cdot \tau \beta \alpha} +
eq_{\tau} {\varphi}^{\lambda}{\sigma}^{\tau}_{\cdot \lambda \beta \alpha} -
q_{\beta} (e{\varphi}^{\tau} {\sigma}^{\lambda}_{\cdot \alpha \tau \lambda} -
q_{\tau} {\varphi}^{\lambda} {\varphi}^{\delta}
{\sigma}^{\tau}_{\cdot \alpha \lambda \delta} ) +$$
$$eq_{\alpha} [{\varphi}^{\lambda} {\sigma}^{\tau}_{\cdot \tau \beta \lambda} -
eq_{\tau} {\varphi}^{\lambda} {\varphi}^{\delta}
{\sigma}^{\tau}_{\cdot \lambda \beta \delta} - {e\over n+m-1}q_{\beta}
( {\varphi}^{\tau} {\varphi}^{\lambda}
{\sigma}^{\delta}_{\cdot \tau \gamma \delta} - eq_{\tau} {\varphi}^{\lambda}
{\varphi}^{\delta} {\varphi}^{\varepsilon}
{\sigma}^{\tau}_{\cdot \lambda \delta \varepsilon} )] \} .$$


Finally, putting the last formula and (B16) into (B15) and after  a rearrangement
of terms we obtain the second group of $na_{(3)}$-invariant conditions (25).
If necessary we can rewrite these conditions in terms of geometrical objects
on $\xi$ and $\underline {\xi}.$ To do this we mast introduce splittings
(B14) into (25).
\end{enumerate}


\begin{references}

\bibitem{1}  R.\ Miron and M.\ Anastasiei, {\it The Geometry of Lagrange Spaces: Theory and
Applications} (Kluwer, Dordrecht, 1994).
\bibitem{2} R.\ Miron and M.\ Anastasiei, {\it Vector Bundles. Lagrange Spaces.
Applications in Relativity} (Academiei, Romania, 1987), (in Romanian).
\bibitem{3} {\it Mathematical and Computer Modelling,} {\bf 20,} N 415, edited by
A.\ Antonelli and T.\ Zastavneak (Plenum Press, 1994).
\bibitem{4} {\it Lagrange and Finsler Geometry, Applications to Physics and Biology},
edited Peter L.\ Antonelli and Radu Miron (Kluwer, Dordrecht, 1996).
\bibitem{5} G.\ S.\ Asanov, {\it Finsler Geometry, Relativity and Gauge Theories}
(Reidel, Boston, 1985).
\bibitem{6} M.\ Matsumoto {\it Foundations of Finsler Geometry and Special Finsler Spaces}
(Kaisisha,Shigaken, 1986).
\bibitem{7} H.\ Rund {\it The Differential Geometry of Finsler Spaces} (Springer--Verlag,
Berlin, 1959).
\bibitem{8} A.\ Bejancu {\it Finsler Geometry and Applications} (Ellis Horwood, Chichester,
England, 1990).
\bibitem{9} P.\ Finsler {\it $\ddot U$ber Kurven und Fl$\ddot a$chen in Allgemeiner
R$\ddot a$men}(Dissertation, G$\ddot o$ttingen, 1918); reprinted
(Birkh$\ddot a$user, Basel, 1951).
\bibitem{10} E.\ Cartan {\it Expos$\acute e$s de G$\acute e$om$\acute e$trie} in {\em Series
Actualit$\acute e$s Scientifiques et Industrielles,}{\bf 79;} reprinted (Herman,
Paris, 1971).
\bibitem{11} S.\ Vacaru, J.\ Math.\ Phys.\ {\bf 37,} 508 (1996).
\bibitem{12} S.\ Vacaru and Yu.\ Goncharenko, Int.\ J.\ Theor.\ Phys.\ 
 {\bf 34,} 1955 (1995).
\bibitem{13} M.\ Anastasiei and H.\ Kawaguchi, Tensor, N.\ S.\ , {\bf 49,}
296 (1990).
\bibitem{14} A.\ Mo\'or,\  Acta Math.\ {\bf 86,} 71 (1951).
\bibitem{15} J.\ Gottlieb, V.\ Oproiu and G.\ Zet , An.\  \c{S}ti.\  Univ.\
"Al.\ I.\ Cuza" Iasi,\ Sect.\ Ia.\ Mat.\ ( N.\ S.\ ) {\bf 20,} 123 (1974).
\bibitem{16} S.\  Vacaru and S.\  Ostaf, Buletinul Academiei de
\c{S}tiin\c{t}e a Republicii Moldova, Fizica \c{s}i  Tehnica, {\bf 1(13),} 64 
 (1994).
\bibitem{17} S.\ Vacaru, S.\ Ostaf and Yu.\ Goncharenko,\ Romanian Journal
of Physics, {\bf 39,} 199 (1994).
\bibitem{18} S.\ Vacaru and S.\ Ostaf,  Buletinul Academiei de
 \c{S}tiin\c{t}e a Republicii Moldova, Fizica \c{s}i tehnica, {\bf 3,}
4  (1993).
 \bibitem{19} S.\ Vacaru, in {\it Contr. Int. Conf. " Lobachevski and
 Modern Geometry ", Part II, } edited by V. Bajanov et all (Kazani,
 University Press,    1992) 64.
\bibitem{20}  S.\ Vacaru, Buletinul Academiei de \c{S}tiin\c{t}e a Republicii
    Moldova, Fizica \c{s}i Tehnica,  {\bf 1(13),} 72 (1994).
\bibitem{21} S.\ Vacaru,  Buletinul Academiei de \c{S}tiin\c{t}e a ,Republicii
    Moldova, Fizica \c{s}i Tehnica,  {\bf 3,} 17 (1993).
\bibitem{22} N.\ S.\ Sinyukov : {\it Geodesic Maps of Riemannian Spaces ,}
 (Nauka, Moscow, 1979) (in Russian).
\bibitem{23} J.\ Kern, Arch. Math., {\bf 25,} 438 (1974).
\bibitem{24} S.\ Vacaru, S.\ Ostaf, Yu.\ Goncharenko and A. Doina,
      Buletinul Academiei de \c{S}tiin\c{t}e a Republicii
     Moldova, Fizica \c{s}i tehnica,  {\bf 3(15),} 42 (1994).
\bibitem{25} S.\ Vacaru and S.\ Ostaf, in {\it Lagrange and Finsler Geometry,
 Applications to Physics and Biology}, edited by Peter L.\ Antonelli 
 and Radu Miron (Kluwer, Dordrecht, 1996).
\bibitem{26} S.\ Vacaru and S.\ Ostaf, in {\it Colloquium on Differential
Geometry, 25-30 July, 1994, Debrecen, Hungary} (Lajos Kossuth University,
Debrecen, Hungary, 1994), 56.
\bibitem{27} I.\ Gottlieb and S.\ Vacaru, in {\it Colloquium on Differential
Geometry, 25-30 July, 1994, Debrecen, Hungary} (Lajos Kossuth University,
Debrecen, Hungary, 1994), 9.
\bibitem{28} S.\ Vacaru,  Buletinul Academiei de \c{S}tiin\c{t}e a Republicii
    Moldova, Fizica \c{s}i Tehnica,  {\bf 2,}  (1995).
\bibitem{29} I.\ Gottlieb and S.\ Vacaru, in {\it Lagrange and Finsler Geometry,
 Applications to Physics and Biology}, edited by  Peter L.\ Antonelli 
 and Radu Miron (Kluwer, Dordrecht, 1996).
\bibitem{30} S.\ Vacaru, Romanian Journal of Physics, {\bf 39,} 37 (1994).
\bibitem{31} E.\ Cartan {\it  Les Systems Differentielles Exterieurs et Lewrs
     Application Geometricques, } (Herman and Cie Editeur, Paris, 1945).
\bibitem{32} J.\ L.\ Synge {\it Relativity: The General Theory,}
(North-Holland Publishing Company, Amsterdam, 1960).
\bibitem{33} B.\ S.\ DeWitt and R.\ W.\ Brehme, Ann.\ Phys.\ {\bf 9,} 220 (1960).
\bibitem{34} B.\ S.\ DeWitt {\it Dynamical Theory of Groups and Fields,}
(Gordon and Breach,  New York, 1965).
\bibitem{35} N.\ Rosen, Phys. Rev. {\bf 57,} 147 (1940).
\bibitem{36} Z.\ Kohler, Z. Physik, {\bf 134,} 286 (1954).
\bibitem{37} N.\ A.\ Chernikov, Preprint JINR P2-88-27, Dubna, 1988
(in Russian).
\bibitem{38} L.\ P.\ Grishchuk, A.\ N.\ Petrov and A.\ D.\ Popova, Commun.\
 Math.\ Phys.\ {\bf 94,} 379 (1984).
\bibitem{39} A.\ A.\ Logunov and M.\ A.\ Mestvirishvily {\it Relativistic
Theory of Gravitation} (Nauka, Moscow, 1989) (in Russian).
\bibitem{40} I.\ Gottlieb and N.\ Ionescu--Pallas, Rev.\ Roum.\ Phys.,\ {\bf 35,}
 25 (1990).
\end{references}
\end{document}